%% file: draftJHEP.tex
\documentclass[a4paper,11pt,final]{article}
\pdfoutput=1 

\usepackage{jheppub} 

\usepackage[T1]{fontenc} 

\usepackage[bordercolor=black,backgroundcolor=yellow,linecolor=black,textsize=scriptsize,colorinlistoftodos,obeyFinal]{todonotes}

\newcommand{\Comment}[2][backgroundcolor=gray!30!white]{ \todo[backgroundcolor=gray!30!white,bordercolor=gray!30!white,#1]{#2}}

\renewcommand{\Re}{\text{Re}}

\newcommand{\mA}{\mathcal{A}}
\newcommand{\mB}{\mathcal{B}}
\newcommand{\mC}{\mathcal{C}}
\newcommand{\mD}{\mathcal{D}}
\newcommand{\mE}{\mathcal{E}}

\newcommand{\mL}{\mathcal{L}}

\newcommand{\mR}{\mathcal{R}}

\newcommand{\mO}{\mathcal{O}}

\newcommand{\comm}[1]{\textcolor{red}{#1}}

\newcommand{\mV}{\mathcal{V}}

\newcommand{\U}{U(\varepsilon)}
\newcommand{\e}{\varepsilon}
\newcommand{\Sn}{\sum_{n}}

\newcommand{\Smn}{\sum_{m,n}}

\usepackage{bbm}
\usepackage{physics}

\newcommand{\Poincare}{Poincar\'e}

\newcommand{\PC}{Poincar\'{e}}

\newcommand{\ep}{\sigma}

\newcommand{\qed}{\nobreak \ifvmode \relax \else\ifdim\lastskip<1.5em
\hskip-\lastskip\hskip1.5em plus0em minus0.5em \fi \nobreak\vrule height0.75em
width0.5em depth0.25em\fi}

\newcommand{\be}{\begin{eqnarray}}
\newcommand{\ee}{\end{eqnarray}}

\def\>{\rangle}
\def\<{\langle}

\newcommand{\executeiffilenewer}[3]{%
\ifnum\pdfstrcmp{\pdffilemoddate{#1}}%
{\pdffilemoddate{#2}}>0%
{\immediate\write18{#3}}\fi%
}

\usepackage{subcaption}
\graphicspath{{plots/}}
\newcommand{\pre}{Phys. Rev. E}

\usepackage{hyperref}
\hypersetup{
	pdftitle    = {Complexity change under conformal transformations in AdS3/CFT2},
	pdfauthor   = {Mario Flory, Nina Miekley}
}

\title{\boldmath Complexity change under conformal transformations in $AdS_{3}/CFT_{2}$}

\author[a,1]{Mario Flory,\note{Corresponding author.}}
\author[b]{and Nina Miekley}

\affiliation[a]{Institute of Physics, Jagiellonian University, \\
\L{}ojasiewicza 11, 30-348 Krak\'ow, Poland}
\affiliation[b]{Lehrstuhl f\"ur Theoretische Physik III, Institut
	f\"ur Theoretische Physik und Astrophysik, \\ Julius-Maximilians-Universit\"at W\"urzburg,
	Am Hubland, D-97074 W\"urzburg, Germany}

\emailAdd{mflory@th.if.uj.edu.pl}
\emailAdd{nina.miekley@physik.uni-wuerzburg.de}

\abstract{Using the volume proposal, we compute the change of complexity of holographic states caused by a small conformal transformation in $AdS_{3}/CFT_{2}$. This computation is done perturbatively to second order. We give a general result and discuss some of its properties. As operators generating such conformal transformations can be explicitly constructed in CFT terms, these results allow for a 
comparison between holographic methods of defining and computing computational complexity and purely field-theoretic proposals. A comparison of our results to one such proposal is given.}

\begin{document}
\maketitle
\flushbottom

\setcounter{tocdepth}{1}
\thispagestyle{empty}

	\input{Intro}

	\input{SGD}
	\input{CV}
	\input{NinasResults}

\input{axiomatic}
	\input{Conc}

\acknowledgments

We would like to thank the following people for useful discussions: Dmitry Ageev, Martin Ammon, Andrey Bagrov, Ralph Blumenhagen, Pawel Caputa,
Johanna Erdmenger, Federico Galli, Micha\l{} Heller,
Haye Hinrichsen, Romuald Janik, Keun-Young Kim, Javier Mag\'{a}n, Charles Melby-Thompson, Ren\'{e} Meyer, Max Riegler, \'Alvaro V\'eliz-Osorio and Runqiu Yang.

MF was supported by the Polish National Science Centre (NCN) grant 2017/24/C/ST2/00469. We would also like to thank the Max Planck Institute for Physics and the Galileo Galilei Institute for Theoretical Physics for the hospitality and the INFN for partial support during the completion of this work.

This research was supported in part by Perimeter Institute for Theoretical Physics. Research at
Perimeter Institute is supported by the Government of Canada through the Department of
Innovation, Science and Economic Development and by the Province of Ontario through the
Ministry of Research, Innovation and Science.

\appendix

\input{AppendixCFT2New}


\providecommand{\href}[2]{#2}\begingroup\raggedright\endgroup

\end{document}

%% file: Intro.tex
\section{Introduction}
\label{sec::Intro} 

Suppose that a scientist in possession of a quantum computer is given a specific task like, for example, applying a certain operator $U$ to an initial reference state $\left|\mR\right\rangle$ in order to obtain the resulting state 
\begin{align}
\left|\psi_U\right\rangle=U\left|\mR\right\rangle. \label{eqn:DefU}
\end{align}
In \cite{2005quant.ph..2070N,2006Sci...311.1133N} it was proposed to define the \textit{quantum computational complexity} (short \textit{complexity} from here on) of the operator $U$ by geometric methods, defining a distance measure on the space of unitary operators and equating the complexity of $U$, $\mC(U)$, as the (minimal) distance between $U$ and the identity operator $\mathbbm{1}$ according to this distance function. 
Equivalently, the complexity of $\left|\psi_U\right\rangle$ with respect to $\left|\mR\right\rangle$, $\mC(\left|\psi_U\right\rangle,\left|\mR\right\rangle)$, could be defined to be the minimal complexity of any operator $U'$ such that $\left|\psi_U\right\rangle=U'\left|\mR\right\rangle$.\footnote{\label{f1} $U'$ might be equal to $U$, or it may be a more efficient operator in terms of complexity. Hence $\mC(U)\geq\mC(U\left|\psi\right\rangle,\left|\psi\right\rangle)$ in general.}
The idea behind this is that in order to implement the operation, the programmer of the quantum computer would have to subdivide $U$ into a product of allowed universal gates that implement the operation step by step, until the end-state agrees with the desired outcome $\left|\psi_U\right\rangle$ at least within a certain error tolerance. This notion of complexity is meant to count the minimal number of gates that the programmer would have to utilise even when using an optimal program. This definition of the complexity
would hence neccessarily depend on the following explicit or implicit choices (see also \cite{Chapman:2017rqy,Agon:2018zso}):
\begin{itemize}
\item A choice of the \textit{reference state} $\left|\mR\right\rangle$, which often is assumed to be a simple product state, without spacial entanglement. This should \textit{not} be confused with the groundstate $\left|0\right\rangle$ of a given physical system.
\item A choice of the \textit{set of allowed gates} $\{\mu_i\}$, such that the operation can be decomposed as 	$U \approx \mu_1 \mu_2 \mu_3 ...$
\item ...within a specified \textit{error tolerance} 
(in some distance measure between $U$ and $\mu_1 \mu_2 \mu_3$).
\end{itemize}

Starting with \cite{Harlow:2013tf,Susskind:2013aaa,Susskind:2014rva}, the idea of computational complexity has begun to see rising interest from the AdS/CFT community. A tentative entry to the holographic dictionary called the \textit{volume proposal} was proposed and motivated in \cite{Susskind:2013aaa,Susskind:2014rva,Susskind:2014moa,Stanford:2014jda,Susskind:2014jwa}.
According to this, the complexity $\mC$ of a field theory state with a smooth holographic dual geometry should be measured by the volumes $\mV$ of certain spacelike extremal co-dimension one bulk hypersurfaces, i.e.   
\begin{align}
\mC\propto \frac{\mV}{LG_N},
\label{complexity}
\end{align} 
wherein a length scale $L$ has to be introduced into equation \eqref{complexity} for dimensional reasons which is usually picked to be the AdS scale \cite{Stanford:2014jda,Susskind:2014jwa,Brown:2015bva,Brown:2015lvg}. 

Interestingly, computational complexity is not the only field theory quantity that has been proposed to be holographically dual to the volumes of extremal co-dimension one hypersurfaces in the bulk. In \cite{MIyaji:2015mia}, it was argued that the volume $\mV$ of an extremal spacelike co-dimension one hypersurface is approximately dual to a quantity $G_{\lambda\lambda}$ called \textit{fidelity susceptibility} \cite{2007PhRvE..76b2101Y,2015JPCM...27t5601Y} according to the formula
\begin{align}
G_{\lambda\lambda}=n_d \frac{\mV}{L^{d}},
\label{fidsus}
\end{align}
where $n_d$ is an order one factor, $L$ is the AdS radius and $d$ determines the dimension such that the AdS space is $d+1$ dimensional. Given two normalised states $\left|\psi(\lambda)\right>$ and $\left|\psi(\lambda+\delta\lambda)\right>$ depending on one parameter $\lambda$, $G_{\lambda\lambda}$ is defined as\footnote{The name fidelity susceptibility derives from the fact that $|\left<\psi(\lambda)\middle|\psi(\lambda+\delta\lambda)\right>|$ is called the fidelity.}
\begin{align}
|\left<\psi(\lambda)\middle|\psi(\lambda+\delta\lambda)\right>|=1-G_{\lambda\lambda}\delta\lambda^2+\mO(\delta\lambda^3)
\label{fidsusdef}
\end{align}
and measures the distance between the two states in a sense, hence it may also be referred to as the \textit{quantum information metric} \cite{Braunstein:1994zz}. The derivation of \eqref{fidsus} in \cite{MIyaji:2015mia} (see also \cite{Bak:2015jxd,Trivella:2016brw}) assumed that two states $\left|\psi(\lambda)\right>$ and $\left|\psi(\lambda+\delta\lambda)\right>$ are the ground states of a theory allowing for a holographic dual, and that the difference $\delta\lambda$ is the result of a perturbation of the Hamiltonian by $\delta\lambda \cdot\hat{O}$ with an exactly marginal operator $\hat{O}$. The bulk spacetime dual to this field theory problem is a \textit{Janus solution} \cite{Bak:2003jk,Bak:2007jm}, but as shown in \cite{MIyaji:2015mia} this geometry can be approximated by a simpler spacetime with a probe defect brane embedded into it, leading to \eqref{fidsus}. This proposal has been utilised for holographic calculations in \cite{Momeni:2016qfv,Sinamuli:2016rms,Banerjee:2017qti,Flory:2017ftd}, 
and our results concerning changes $\delta\mV$ induced by infinitesimal conformal transformations, to be derived in section \ref{sec::CV}, may also have an interesting physical interpretation from the perspective of fidelity susceptibility, however in this paper we will focus on bulk volumes as a holographic dual to computational complexity.

The necessity to include a lengthscale in the definition \eqref{complexity} of holographic complexity was considered unsatisfactory by some, and so \cite{Brown:2015bva,Brown:2015lvg} proposed the competing \textit{action proposal}
\begin{align}
\mC=\frac{\mA}{\pi\hbar}
\label{action}
\end{align}
wherein $\mA$ is the bulk action over a certain (co-dimension zero) bulk region, the Wheeler-DeWitt patch.

Both the volume- and action proposals for holographic complexity where subsequently used for a number of holographic investigations. Specific topics of interest where the boundary terms required to calculate the action $\mA$ in \eqref{action} correctly \cite{Lehner:2016vdi}, the time dependence of complexity \cite{Yang:2016awy,Brown:2017jil,Carmi:2017jqz,Moosa:2017yvt,Kim:2017qrq,Ageev:2018nye} (especially with respect to the so called Lloyd's bound, see however \cite{Cottrell:2017ayj,Moosa:2017yiz}), generalisations of holographic complexity to mixed states \cite{Alishahiha:2015rta,Ben-Ami:2016qex,Carmi:2016wjl,Abt:2018ywl,Agon:2018zso}, RG-flows \cite{Flory:2017ftd,Roy:2017uar}, three-dimensional gravity models \cite{Ghodrati:2017roz,Qaemmaqami:2017lzs} and many others. Interesting connections have been made between holographic complexity and kinematic space approaches \cite{Abt:2017pmf,Abt:2018ywl} as well as Liouville theory in two dimensions \cite{Caputa:2017urj,Caputa:2017yrh,Czech:2017ryf,Bhattacharyya:2018wym}.\footnote{Although a comparably young topic, the literature on holographic complexity has indeed grown to a considerable size by now. We apologize to everyone who feels they where unjustly left out above. }

This amount of progress on the holographic side has also led to increased efforts to provide better and more concrete definitions of quantum computational complexity in quantum mechanical or even quantum field theory contexts \cite{Jefferson:2017sdb,Chapman:2017rqy,Caputa:2017urj,Caputa:2017yrh,Yang:2018nda,Hashimoto:2018bmb,Reynolds:2017jfs,Yang:2017nfn,Magan:2018nmu,Khan:2018rzm,Kim:2017qrq}.
However, in the proposals \eqref{complexity} and \eqref{action} for holographic complexity calculations it is not clear what the relevant reference state, gate set and error tolerance are to be. If one or both of these proposals for holographic complexity are to be correct, then these choices need to be somehow \textit{implicit} in the holographic dictionary.

This is the main motivation of our present work: We will, focusing on the volume proposal \eqref{complexity} for now, calculate how the complexity of a given state of a two dimensional conformal field theory (CFT$_2$) with smooth holographic dual changes under a conformal transformation. Such conformal transformations can of course be applied to any two dimensional CFT, irrespectively of whether the central charge is large or not, or whether the CFT satisfies any other requirement for having a holographic dual.
Hence we believe these results will be useful in the future for comparisons between the holographic and field theory proposals for complexity, and this may help to elucidate the choices of gateset, reference state and error tolerance that are implicit in the holographic proposals.\footnote{
See also \cite{Reynolds:2017jfs,Fu:2018kcp} for recent papers comparing results from the action- and volume-proposal and discussing what they may imply for a possible field theory definition of complexity.}

The structure of our paper is as follows: In section \ref{sec::SGD}, we will review how to implement a (small) conformal transformation in AdS$_3$/CFT$_2$. This will provide the general setup of our work and fix the notation. Then, based on the volume proposal \eqref{complexity}, we will calculate in section \ref{sec::CV} how holographic complexity changes under small conformal transformations. A few illustrative examples will be discussed in section \ref{sec::Examples}, and in section \ref{sec::FTproposal} we will investigate which constraints our results imply on the reference state $\left|\mR\right\rangle$ in the framework of a field theory proposal for complexity made in \cite{Yang:2018nda}.
We close in section \ref{sec::Conc} with a conclusion and outlook. Some technical details about the implementation of conformal transformations on the field theory side will be presented in appendix \ref{sec::AppCFT2}.

Note: While in the final stages of preparing this draft, we became aware of the upcoming paper \cite{Caputa:2018kdj} that seems to share some common ideas with ours, however approaching the topic from the field theory side.

%% file: SGD.tex
\section{Gravity setup}
\label{sec::SGD}

We consider the vacuum state of a two-dimensional CFT with a classical holographic dual. The bulk geometry is given by AdS$_3$ space which in the \Poincare-patch is written as
\begin{align}
	ds^2 =& \frac{d\lambda^2}{4 \lambda^2} - \lambda dx^+\cdot dx^-,
	\label{eqn:CoordForSGD}
\end{align}
where $\lambda$ is our radial coordinate and $x^\pm = t \pm x$ ($-\infty<t,x<+\infty$) are the light-cone coordinates of the field theory. In these coordinates, the boundary of AdS is at $\lambda = \infty$ and the \PC{} horizon at $\lambda = 0$. Placing a cutoff at $\lambda=1/\epsilon^2$ ($\epsilon\rightarrow0$), it is possible to holographically calculate the expectation value of the energy-momentum tensor of the boundary CFT by the method of \cite{Balasubramanian:1999re}, which gives a vanishing result as expected.

Although Einstein-Hilbert gravity in three dimensions is trivial in the sense of having no propoagating degrees of freedom, there is a surprising number of vacuum solutions which were derived in \cite{Banados:1998gg} (see also \cite{Mandal:2014wfa}). The reason is that these solutions, called \textit{Ba\~{n}ados geometries}, can be derived from \eqref{eqn:CoordForSGD} by diffeomorphisms which are
\textit{global} in the sense that they act nontrivially near the boundary, such that while the resulting metric is still asymptotically AdS in the appropriate sense, the holographic energy-momentum tensor has changed. Despite all being locally isometric to AdS$_3$ in the bulk, these geometries are hence dual to \textit{different states} of the dual field theory.\footnote{These states are generically time dependent, and can for example be used for studies of thermalisation or equilibration processes \cite{Bhaseen:2013ypa,Erdmenger:2017gdk}. See also \cite{Sheikh-Jabbari:2014nya,Sheikh-Jabbari:2016unm,Sheikh-Jabbari:2016znt} for more detailed studies of such Ba\~{n}ados geometries.}
In \cite{Mandal:2014wfa}, these transformations were hence termed \textit{solution generating diffeomorphisms (SGDs)}, and in this section we will explain these transformations in detail, following the outline and notation of \cite{Mandal:2014wfa}.

\begin{figure}[htb]
	\centering
	\def\svgwidth{0.4\columnwidth}
\executeiffilenewer{ConformalDiagram.svg}{ConformalDiagram.pdf}%
{inkscape -z -D --file=ConformalDiagram.svg %
--export-pdf=ConformalDiagram.pdf --export-latex}%
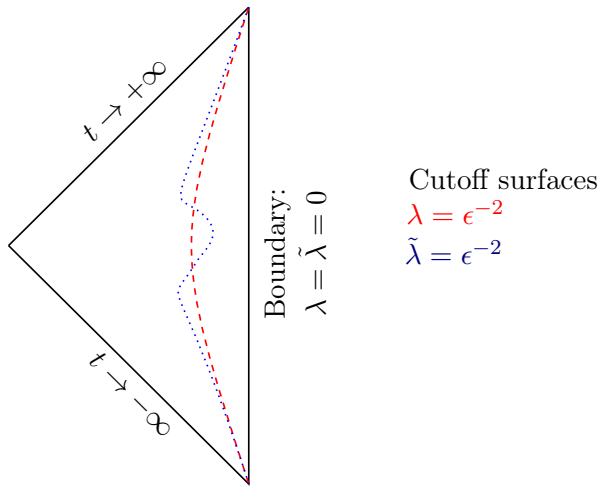%

	\caption{A conformal diagram of the \Poincare-patch of AdS$_3$. The vertical line is the asymptotic boundary while the two diagonal lines are the two \Poincare-horizons where $t\rightarrow\pm\infty$. The two cutoff surfaces $\lambda=1/\epsilon^2$ and $\tilde{\lambda}=1/\epsilon^2$ are shown as dashed (red) and dotted (blue) lines, respectively.}
	\label{fig:CutOffSurfaces}
\end{figure}

In order to apply a SGD to the geometry described by \eqref{eqn:CoordForSGD}, we perform a coordinate transformation\footnote{One virtue of the SGDs formulated in \cite{Banados:1998gg} was that they preserved a specific form of the metric tensor. As can be seen from equation \eqref{metricafterSGD}, this is not the case for the SGDs as used in \cite{Mandal:2014wfa} and herein. As can be read-off from the holographic energy-momentum tensor \eqref{T}, both types of SGD implement conformal transformations on the CFT-state.
The apparent difference between the SGDs comes merely from the fact that the SGDs of \cite{Banados:1998gg} and the corresponding SGDs of \cite{Mandal:2014wfa} differ by a bulk-diffeomorphism that acts trivially at the boundary. We mostly follow the convention and notation of \cite{Mandal:2014wfa} as therein the authors also explicitly discuss how to implement such SGDs on two-sided black holes and thereby construct an infinite family of purifications of, e.g., the thermal state. We hope to investigate this class of solutions from the viewpoint of holographic complexity in the future.}
\begin{subequations}
	\begin{align}
	\lambda &= \frac{\tilde \lambda}{G'_+(\tilde x^+) G'_-(\tilde x^-)}, \\
	x^+ &= G_+(\tilde x^+), \\
	x^- &= G_-(\tilde x^-).
	\end{align}\label{eqn:SGD}%
\end{subequations}
The metric in the tilted coordinates is \cite{Mandal:2014wfa}
\begin{subequations}
\begin{align}
  ds^2 &= \frac{1}{4\tilde \lambda^2} d\tilde\lambda^2 -\tilde \lambda~ d\tilde x^+\cdot d\tilde x^-+\left( A_+ d\tilde x^+ + A_- d\tilde x^-\right)^2 + \frac{1}{\tilde \lambda}   d\tilde \lambda \cdot \left(A_+ d\tilde x^+ +A_- d\tilde x^-  \right), \\
  A_\pm &= -\frac{1}{2}\frac{G_\pm{}''(\tilde x^\pm)}{G_\pm{}'(\tilde x^\pm)}.
\end{align}
\label{metricafterSGD}%
\end{subequations}
In principle, general relativity is invariant under coordinate transformations. However, these SGDs fall off slowly at the boundary and are asymptotically non-trivial gauge transformation. The new cutoff is at
\begin{align}
  \tilde \lambda &= \frac{1}{\epsilon^2} \Leftrightarrow
  \lambda = \frac{1}{\epsilon^2G_+{}'(\tilde{x}^+) G_-{}'(\tilde{x}^-)},
  \label{cutoff}
\end{align}
with the field theory UV-cutoff $\epsilon$. This shows the non-triviality of the coordinate transformation: in terms of the old coordinates, the cutoff surface is wrapped and different to the one which describes the vacuum state (i.e.~$\lambda = 1/\epsilon^2$). This is shown in Figure \ref{fig:CutOffSurfaces}.

We started with a CFT state with vanishing energy-momentum tensor. In the deformed state, with the cutoff defined as in \eqref{cutoff}, we have \cite{Mandal:2014wfa}
\begin{subequations}
\begin{align}
  8 \pi G_3 T_{++ } &= \frac{1}{4G_+{}'(\tilde{x}^+)^2}\left(3G_+{}''(\tilde{x}^+)^2-2G_+{}'(\tilde{x}^+)G_+{}'''(\tilde{x}^+) \right), \\
  8 \pi G_3 T_{-- } &= \frac{1}{4G_-{}'(\tilde{x}^-)^2}\left(3G_-{}''(\tilde{x}^-)^2-2G_-{}'(\tilde{x}^-)G_-{}'''(\tilde{x}^-) \right), \\
  8 \pi G_3 T_{+- } &=0,
\end{align}
  \label{T}%
\end{subequations}
which is precisely what we expect for the change of the energy-momentum tensor after applying a conformal transformation to the groundstate, see appendix \ref{sec::AppCFT2}. The SGDs hence implement conformal transformations on the boundary theory, which is also evident from the change of the induced metric of the cutoff surface under these transformations. In field theory terms, the conformal transformation with functions $G_{\pm}$ hence maps the ground state $\left|0\right\rangle$ to a new state
\begin{align}
\left|\psi(G_{+},G_{-})\right\rangle=U(G_+)U(G_-)\left|0\right\rangle
\label{Uconf}
\end{align}
with known (commuting) unitary operators $U(G_+)$ and $U(G_-)$, see \cite{Mandal:2014wfa} and appendix \ref{sec::AppCFT2} for an explicit construction in the case of a small conformal transformation.

In the following, we consider a small SGD, i.e.~
\begin{subequations}
	\begin{align}
	x^+=G_+(\tilde x^+) &= \tilde x^+ +\ep ~ g_{+}(\tilde x^+) ,  \\
	x^-=G_-(\tilde x^-) &= \tilde x^- +\ep ~  g_{-}(\tilde x^-),
	\end{align}
	\label{SmallSGD}%
\end{subequations}
with the expansion parameter $\ep \ll 1$. The metric defined by the line element \eqref{metricafterSGD} can then be expanded in orders of $\ep$ and we find
\begin{align}
  ds^2 &= \frac{d\tilde \lambda^2}{4 \tilde\lambda^2} - \tilde\lambda d\tilde t^2 + \tilde\lambda d\tilde x^2 \nonumber \\
  &~~~-\frac{\ep}{2 \tilde\lambda} \Big[ \Big(g_{+}''\left(\tilde{t}+\tilde{x}\right)+g_{-}''\left(\tilde{t}-\tilde{x}\right)\Big)d\tilde t
  +\Big(g_{+}''\left(\tilde{t}+\tilde{x}\right)-g_{-}''\left(\tilde{t}-\tilde{x}\right)\Big)d\tilde x\Big]d\tilde \lambda +\mO(\ep^2)
  \label{metric}
\end{align}
where we have switched from lightcone coordinates $\tilde{x}_\pm$ to standard coordinates $\tilde{t},\tilde{x}$ on the boundary.
The components of the energy-momentum tensor read
\begin{subequations}
	\begin{align}
	8 \pi G_3 T_{++ } &=- \frac{\ep}{2} g_+{}'''(\tilde x^+)+ \frac{\ep^2}{4}\left(3g_+{}''(\tilde x^+)^2+2g_+{}'(\tilde x^+)g_+{}'''(\tilde x^+) \right) + \mO(\ep^3), \\
	8 \pi G_3 T_{-- } &= -\frac{\ep}{2} g_-{}'''(\tilde x^-)+ \frac{\ep^2}{4}\left(3g_-{}''(\tilde x^+)^2+2g_-{}'(\tilde x^+)g_+{}'''(\tilde x^-) \right) + \mO(\ep^3), \\
	8 \pi G_3 T_{+- } &=0.
	\end{align}
	\label{Ttosecondorder}%
\end{subequations}
For later, it will be convenient to introduce the Fourier transforms of the functions $g_\pm$:
\begin{subequations}
	\begin{align}
		g_+(\tilde x^+)&= \int\limits_{-\infty}^{\infty}  d\xi  ~\hat g_+ ( \xi) \exp (-2 \pi i\cdot \tilde x^+ \cdot \xi)= \int\limits_{-\infty}^{\infty} d\xi ~\hat g_{+}(-\xi) e^{2\pi i \xi \tilde{t}}~ \exp(+2\pi i \xi\cdot   \tilde{x})
,\\
		g_-(\tilde x^-)&=  \int\limits_{-\infty}^{\infty}  d\xi  ~\hat g_- ( \xi) \exp (-2 \pi i\cdot \tilde x^- \cdot \xi)= \int\limits_{-\infty}^{\infty} d\xi ~\hat g_{-}(-\xi) e^{2\pi i \xi \tilde{t}}~ \exp(-2\pi i \xi\cdot   \tilde{x}).
	\end{align}\label{Fourierg}%
\end{subequations}
In the following, we will always assume that $g_\pm$ falls off to zero towards infinity and are bounded. As a consequence of $g_\pm$ being real, the Fourier transforms have to satisfy
\begin{align}
	\hat g_\pm^*(\xi) = \hat g_\pm(-\xi). \label{eqn:RestrictionFourierTransform}
\end{align}

\Comment[caption={Expansion of metric },inline]{
	\begin{minipage}{0.825\linewidth}
		alternative form:
		\begin{align}
		ds^2 &= \frac{d\tilde \lambda^2}{4 \tilde\lambda^2} - \tilde\lambda d\tilde x^+\cdot d\tilde x^-
		-\frac{\ep}{2 \tilde\lambda} \Big[ g_{-}''\left(\tilde{x}^-\right)d\tilde x^-
		+g_{+}''\left(\tilde{x}^+\right)d\tilde x^+ \Big]d\tilde \lambda +\mO(\ep^2).
		\end{align}
	\end{minipage}
}
\Comment[caption={Expansion of metric  II},inline]{
	\begin{minipage}{0.825\linewidth}
		\begin{subequations}
			\begin{align}
			g_{\tilde t \tilde t} &= -\tilde{\lambda }+\frac{1}{4} \ep  ^2 \left(g_{-}''\left(\tilde{t}-\tilde{x}\right){}^2+2 g_{+}''\left(\tilde{t}+\tilde{x}\right) g_{-}''\left(\tilde{t}-\tilde{x}\right)+g_{+}''\left(\tilde{t}+\tilde{x}\right){}^2\right)+\mO \left(\ep  ^3\right) \\
			g_{\tilde t \tilde x} &= \frac{1}{4} \ep  ^2 \left(g_{+}''\left(\tilde{t}+\tilde{x}\right){}^2-g_{-}''\left(\tilde{t}-\tilde{x}\right){}^2\right)+\mO \left(\ep  ^3\right) \\
			g_{\tilde t \tilde \lambda } &= -\frac{\ep }{4 \tilde{\lambda }}  \left(g_{-}''\left(\tilde{t}-\tilde{x}\right)+g_{+}''\left(\tilde{t}+\tilde{x}\right)\right) \\
			&~+\frac{\ep  ^2}{4 \tilde{\lambda }} \left(g_{-}'\left(\tilde{t}-\tilde{x}\right) g_{-}''\left(\tilde{t}-\tilde{x}\right)+g_{+}'\left(\tilde{t}+\tilde{x}\right) g_{+}''\left(\tilde{t}+\tilde{x}\right)-g_{2,-}''\left(\tilde{t}-\tilde{x}\right)-g_{2,+}''\left(\tilde{t}+\tilde{x}\right)\right)+O\left(\ep  ^3\right) \nonumber\\
			g_{\tilde x \tilde x} &= \tilde{\lambda }+\frac{1}{4} \ep  ^2 \left(g_{-}''\left(\tilde{t}-\tilde{x}\right){}^2-2 g_{+}''\left(\tilde{t}+\tilde{x}\right) g_{-}''\left(\tilde{t}-\tilde{x}\right)+g_{+}''\left(\tilde{t}+\tilde{x}\right){}^2\right)+\mO \left(\ep  ^3\right) \\
			g_{\tilde x \tilde \lambda } &= -\frac{\ep  }{4 \tilde{\lambda }} \left(g_{+}''\left(\tilde{t}+\tilde{x}\right)-g_{-}''\left(\tilde{t}-\tilde{x}\right)\right) \\
			&~ +\frac{\ep  ^2 }{4 \tilde{\lambda }}\left(-g_{-}'\left(\tilde{t}-\tilde{x}\right) g_{-}''\left(\tilde{t}-\tilde{x}\right)+g_{+}'\left(\tilde{t}+\tilde{x}\right) g_{+}''\left(\tilde{t}+\tilde{x}\right)+g_{2,-}''\left(\tilde{t}-\tilde{x}\right)-g_{2,+}''\left(\tilde{t}+\tilde{x}\right)\right)+\mO \left(\ep  ^3\right) \nonumber\\
			g_{\tilde \lambda \tilde \lambda} &= \frac{1}{4 \tilde \lambda^2}
			\end{align}
		\end{subequations}
	\end{minipage}
}

%% file: ConformalDiagram.pdf_tex
\begingroup%
  \makeatletter%
  \providecommand\color[2][]{%
    \errmessage{(Inkscape) Color is used for the text in Inkscape, but the package 'color.sty' is not loaded}%
    \renewcommand\color[2][]{}%
  }%
  \providecommand\transparent[1]{%
    \errmessage{(Inkscape) Transparency is used (non-zero) for the text in Inkscape, but the package 'transparent.sty' is not loaded}%
    \renewcommand\transparent[1]{}%
  }%
  \providecommand\rotatebox[2]{#2}%
  \ifx\svgwidth\undefined%
    \setlength{\unitlength}{315.74998991bp}%
    \ifx\svgscale\undefined%
      \relax%
    \else%
      \setlength{\unitlength}{\unitlength * \real{\svgscale}}%
    \fi%
  \else%
    \setlength{\unitlength}{\svgwidth}%
  \fi%
  \global\let\svgwidth\undefined%
  \global\let\svgscale\undefined%
  \makeatother%
  \begin{picture}(1,1.14251782)%
    \put(1.77873283,1.66545944){\color[rgb]{0,0,0}\makebox(0,0)[lt]{\begin{minipage}{1.70572519\unitlength}\raggedright \end{minipage}}}%
    \put(0,0){\includegraphics[width=\unitlength,page=1]{ConformalDiagram.pdf}}%
    \put(0.2043462,0.7952702){\color[rgb]{0,0,0}\makebox(0,0)[lb]{\smash{\rotatebox{45}{$t\rightarrow+\infty$}}}}    \put(0.91420891,0.61836867){\color[rgb]{1,0,0}\makebox(0,0)[lb]{\smash{$\lambda=\epsilon^{-2}$}}}%
    \put(0.21751806,0.31514185){\color[rgb]{0,0,0}\makebox(0,0)[lb]{\smash{\rotatebox{-45}{$t\rightarrow-\infty$}}}}%
    \put(0,0){\includegraphics[width=\unitlength,page=2]{ConformalDiagram.pdf}}%
    \put(0.91058947,0.53014338){\color[rgb]{0,0,0.50196078}\makebox(0,0)[lb]{\smash{$\tilde\lambda=\epsilon^{-2}$}}}%
    \put(0.60587265,0.40956876){\color[rgb]{0,0,0}\makebox(0,0)[lb]{\smash{\rotatebox{90}{\shortstack{Boundary:\\ $ \lambda = \tilde \lambda =0$}}}}}%
    \put(0.91919752,0.69573386){\color[rgb]{0,0,0}\makebox(0,0)[lb]{\smash{Cutoff surfaces}}}%
  \end{picture}%
\endgroup%

%% file: CV.tex
\section{Complexity = Volume}
\label{sec::CV}


In this section we calculate the complexity of states \eqref{Uconf} for small conformal transformations \eqref{SmallSGD} using the Complexity=Volume (CV) proposal \eqref{complexity}.
For this, we have to calculate the maximal volume of a co-dimension one spacelike slice with fixed boundary conditions, i.e.~we have to maximize
\begin{align}
\mV = \int d \tilde \lambda d \tilde x \sqrt{\gamma}
\label{mV}
\end{align}
with the determinant of the induced metric $\sqrt{\gamma}$ depending on the embedding function $\tilde t(\tilde x,\tilde \lambda)$. This spacelike slice is anchored at the boundary at a constant time slice of the coordinates $\tilde{x}^\pm=\tilde{t}\pm\tilde{x}$.
For a small conformal transformation, we can expand the embedding\footnote{For $\ep=0$, the metric \eqref{metric} is just the \Poincare\ metric, and the appropriate embedding for a maximal volume slice anchored to a constant time-slice on the boundary is just given by $t=t_0$.}
\begin{align}
	\tilde t(\tilde x,\tilde \lambda) =& t_0 + \ep t_1(\tilde x,\tilde \lambda) +\ep^2 t_2(\tilde x,\tilde \lambda)+...\ .
\label{tseries}
\end{align}
Just as the metric functions \eqref{Fourierg}, we can write the
embedding function $t_1$ as an (inverse) Fourier transform
\Comment[caption={Sign convention for $\hat g$},inline]{It is always $\hat g(\xi) e^{-2\pi i t_0 \xi}$ or $\hat g(-\xi) e^{2\pi i t_0 \xi}$. Since the sign convention is different for the embedding, $\hat t(\xi)$ goes together with $\hat g_\pm(\mp \xi)$. }
\begin{align}
t_1(\tilde{x},\tilde{\lambda}) =& \int\limits_{-\infty}^{\infty} d\xi ~ \hat t(\xi,\tilde{\lambda})~ \exp(2\pi i \xi \tilde{x}).
\end{align}
With \eqref{tseries}, we can now expand the integrand of \eqref{mV} in orders of $\ep$, and the lowest nontrivial order leads to equations of motion for $t_1(\tilde{x},\tilde{\lambda})$ of the form
\begin{subequations}
\begin{align}
8 \tilde{\lambda }^3 t_1{}^{(0,2)}\left(\tilde{x},\tilde{\lambda }\right)+20 \tilde{\lambda }^2 t_1{}^{(0,1)}\left(\tilde{x},\tilde{\lambda }\right)+2 t_1{}^{(2,0)}\left(\tilde{x},\tilde{\lambda }\right)&=-g_{+}''\left(\tilde{x}+t_0\right) - g_{-}''\left(t_0-\tilde{x}\right),
\label{RealspaceEOM}
 \\
8 \tilde{\lambda }^3 \hat{t}^{(0,2)}\left(\xi ,\tilde{\lambda }\right)+20 \tilde{\lambda }^2 \hat{t}^{(0,1)}\left(\xi ,\tilde{\lambda }\right)-8 \pi ^2 \xi ^2 \hat{t}\left(\xi ,\tilde{\lambda }\right) &= 4 \pi ^2 \xi ^2 \left( \hat{g}_{+}(-\xi ) e^{2 i \pi  \xi  t_0}+\hat{g}_{-}(\xi ) e^{-2 i \pi  \xi  t_0}\right).
\label{FourierspaceEOM}
\end{align}
\end{subequations}
Therefore, we have a second order partial differential equation for the Fourier coefficients $\hat t$.
The boundary conditions are the fixed behaviour at the asymptotic boundary (i.e.~$\lim_{\tilde \lambda \rightarrow \infty} t_1 = 0$) and at the \Poincare-horizon at $\tilde{\lambda}=0$ we demand that $t_1$ does not diverge to keep the perturbative expansion meaningful. This would mean that the embedding function $t_1$ does not leave the \Poincare-patch through one of the null-segments of the \Poincare-horizon in figure \ref{fig:CutOffSurfaces}, but instead goes into the corner on the left hand side of the figure.\footnote{One can avoid the problem of having to specify boundary conditions at the \Poincare-horizon altogether by mapping the problem to global AdS and solving the equations for the embedding there. Then, it would suffice to give Dirichlet boundary conditions at the full boundary circle of global AdS.}

\Comment[inline,caption={Bdy-term}]{
\begin{minipage}{0.85\linewidth}
	During variation, the boundary term
	\begin{align}
	  0 = \frac{1}{2} \sqrt{\tilde \lambda } \left(-4 \tilde \lambda ^2 t_1{}^{(0,1)}(\tilde x,\tilde\lambda )-g_-''\left(t_0-\tilde x\right)-g_+''\left(t_0+\tilde x\right)\right)
	\end{align}
	has to vanish at $\tilde \lambda = 0$ and $\tilde \lambda = \infty$. We have
	$$ \tilde \lambda ^2 t_1{}^{(0,1)}(\tilde x,\tilde\lambda ) \rightarrow \pi^2 \xi^2 \left(\hat{g}_-(\xi )e^{-2 i \pi  \xi  t_0}+\hat{g}_+(-\xi ) e^{2 i \pi  \xi  t_0}\right)$$
	at the boundary (i.e.\ $\tilde \lambda =\infty$), hence the boundary term at the boundary vanishes ($- 4 \pi^2$ corresponds to the derivatives). The boundary term at the \PC{} horizon vanishes since the embedding does not diverge (too much).
\end{minipage}}

Using these conditions to solve \eqref{FourierspaceEOM}, we obtain a piece-wise smooth result\footnote{In all our calculations, we implicitly assume that that the functions $g_\pm$ fall off to zero towards infinity and are sufficiently well behaved for integrals to be finite and to interchange integration order or integration and differentiation where necessary. Note that for all the specific examples to be discussed in section \ref{sec::Examples}, it is possible to analytically calculate the embedding $t_1(\tilde{x},\tilde{\lambda})$ and confirm that the equation \eqref{RealspaceEOM} as well as the boundary conditions and \eqref{mVresult} are satisfied.}
\begin{align}
\hat t(\xi,\tilde \lambda) =&\left(\frac{1}{2} \left(e^{\frac{-2 \pi  |\xi| }{\sqrt{\tilde \lambda }}}-1\right)+\frac{ \pi  |\xi|}{ \sqrt{\tilde \lambda }} e^{\frac{-2 \pi  |\xi| }{\sqrt{\tilde \lambda }}} \right) \left(\hat{g}_-(\xi )e^{-2 i \pi  \xi  t_0}+\hat{g}_+(-\xi ) e^{2 i \pi  \xi  t_0}\right) .
\end{align}

\Comment[inline,caption={Bdy-term}]{
\begin{minipage}{0.7\linewidth}
I only did the calculation and verified it for one example.
	\begin{align*}
		t_1(x) &= -\frac{1}{2} \Big( g_+(t_0+x)+g_-(t_0-x)\Big) \\ &~~~ + \frac{1}{\pi}\int\limits_{-\infty}^\infty dt \frac{\sqrt{\lambda}}{(\lambda t^2+1)^2} \Big( g_+(t_0+t+x)+g_-(t_0+t-x)\Big)
	\end{align*}
	The limit $\lambda \rightarrow \infty$ has to be taken carefully. With respect to $\lambda$, the integral is an integral transform of the function $g_+(\cdot+t_0+x)+g_-(\cdot+t_0-x)$. The integral kernel approaches $\pi \delta(t)/2$ in the near-boundar limit. \\
	In this form, I see no reason why the embedding should diverge at the Poincare horizon. In contrast, the integral term should just vanish. However, for the Gauss function, this integral couldn't be solved by mathematica. Therefore, I think the divergence is Mathematica going wrong somewhere, but I still think our result should be valid.
\end{minipage}
}

Now, let us turn to the volume. Expanding \eqref{mV} up to second order in $\ep$, we obtain
\footnote{\label{f12}
	The first-order term is
	\begin{align*}
	  \mV_{(1)} = \underbrace{\left.\frac{\delta \mV}{\delta \tilde t}\right|_{\ep=0}}_{0}
	 	  \cdot t_1 +  \underbrace{\left.\frac{\delta \mV}{\delta  g_\pm}\right|_{\ep=0}}_0 \cdot g_\pm =0,
	\end{align*}
	where the first term vanishes because of extremality of the $0th$-order embedding in the \Poincare-metric. It directly follows from this that the second-order term $	\mV_{(2)} $ only depends on the first order terms in the changes of embedding and metric.
}
\begin{align}
  \mV = \left.\mV\right|_{\ep=0 }+  \ep ^2\mV_{(2)} + \mO (\ep^3)
\label{mVseries}
\end{align}
with
\begin{subequations}
\begin{align}
	\mV_{(2)} &= \int\limits_{0}^\infty d \tilde \lambda \int\limits_{-\infty}^\infty d\tilde x \left[-\frac{\sqrt{\tilde{\lambda }}}{2}  \tilde{t}_1{}^{(0,1)}\big(\tilde{x},\tilde{\lambda }\big)\left( g_-''\left(\tilde{t}_0-\tilde{x}\right)+g_+''\left(\tilde{t}_0+\tilde{x}\right)\right)-\tilde{\lambda }^{5/2} \tilde{t}_1{}^{(0,1)}\big(\tilde{x},\tilde{\lambda }\big){}^2-\frac{\tilde{t}_1{}^{(1,0)}\big(\tilde{x},\tilde{\lambda }\big){}^2}{4 \sqrt{\tilde{\lambda }}}\right], \\
	&= \pi ^3\int\limits_{-\infty}^\infty d\xi ~ |\xi| ^3  \left(\hat{g}_+(\xi )e^{-2 i \pi  \xi  t_0}+\hat{g}_-(-\xi ) e^{2 i \pi  \xi  t_0}\right) \left(\hat{g}_-(\xi )e^{-2 i \pi  \xi  t_0}+\hat{g}_+(-\xi ) e^{2 i \pi  \xi  t_0}\right), \label{mVresultb}\\
	&= \pi ^3\int\limits_{-\infty}^\infty d\xi ~ |\xi| ^3  \left|\hat{g}_-(\xi )e^{-2 i \pi  \xi  t_0}+\hat{g}_+(-\xi ) e^{2 i \pi  \xi  t_0}\right|^2 \geq 0,
\end{align}
\label{mVresult}%
\end{subequations}
where we used the properties of the Fourier transforms \eqref{eqn:RestrictionFourierTransform} in the last step.

\Comment[inline,caption={Bdy-term}]{\begin{minipage}{0.75\linewidth}
Combine with results in \cite{Belin:2018bpg}
	\begin{align}
	  \mV_{(2)} =& \frac{3}{4\pi}\int dx dy \frac{1}{\qty(x-y)^4}\Big(g_+(t_0+x)+g_-(t_0-x)\Big) \Big(g_+(t_0+y)+g_-(t_0-y)\Big)
	\end{align}
	\begin{itemize}
		\item $\tilde t = t - \frac{\sigma}{2} (g_+(\tilde x^+)+ g_-(\tilde x^-))$  \\
		$\rightarrow$ interpretation as shift of embedding
	\end{itemize}
\end{minipage}
}


\Comment[inline,caption={Further form for $V_{(2)}$}]{ \begin{minipage}{0.75\linewidth}
Furthermore, we can write $\hat g_\pm$ as
\begin{subequations}
\begin{align}
  \hat{g}_+(\xi) = \hat{g}_+(\xi)~\Theta(\xi) + \hat{g}_+(-\xi)^*~\Theta(-\xi). \\
  \hat{g}_-(\xi) = \hat{g}_-(\xi)~\Theta(\xi) + \hat{g}_-(-\xi)^*~\Theta(-\xi).
\end{align}
\end{subequations}
where $\hat g_\pm(\xi)$ are  completely arbitrary complex functions for $\xi \geq 0$.

\begin{subequations}
	\begin{align}
	  V_{(2),\text{pure}}(\hat{g}) &= 2 \pi ^3\int\limits_{0}^\infty d\xi ~ \xi^3  \cdot \qty |\hat g(\xi) |^2\\
	  V_{(2),\text{mixed}}(\hat{g}_+,\hat{g}_-) &= 2 \pi ^3\int\limits_{0}^\infty d\xi ~ \xi ^3  \cdot\qty( \hat g_+(\xi)\hat g_-(\xi)e^{-4 i \pi \xi t_0}+\hat g_+(\xi)^* \hat g_-(\xi)^* e^{ 4i \pi \xi t_0}) 
	\end{align}
\end{subequations}
\end{minipage}}

\begin{figure}[htb]
	\centering
	\def\svgwidth{0.49\columnwidth}
	\hspace{5cm} %
\executeiffilenewer{ComplexityOfStates2.svg}{ComplexityOfStates2.pdf}%
{inkscape -z -D --file=ComplexityOfStates2.svg %
--export-pdf=ComplexityOfStates2.pdf --export-latex}%
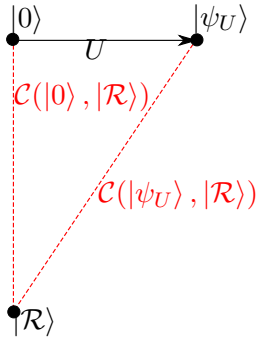%

	\caption[]{The space of states with the reference state $\left|\mR\right\rangle$, the groundstate $\left|0\right\rangle$ and the state $\left|\psi_{U}\right\rangle\equiv U\left|0\right\rangle$ for the generator $U$ of a small conformal transformation. The red dashed lines signify the complexities of the states $\left|\psi_{U}\right\rangle$ and $\left|0\right\rangle$ with respect to the reference state $\left|\mR\right\rangle$.
	}
	\label{fig::ComplexityOfStates}
\end{figure}

We can already point out the following observations: First of all, we see that the change in complexity $\propto\mV_{(2)}$ (see \eqref{complexity}) due to the operators $U$ being applied to the groundstate is always independent of the cutoff $\epsilon$, i.e.~UV finite.\footnote{In the terms of \cite{Chapman:2016hwi}, the \textit{complexity of formation} is finite.} Secondly, in contrast to $V|_{\ep=0}$ which also receives a divergent contribution $\int_{-\infty}^\infty dx=Vol(\mathbb{R})$, this result will be finite due to our assumptions on the boundedness and the falloff conditions at infinity of $g_\pm$.
Thirdly, we find
\begin{align}
\mC(U(g_+)U(g_-)\left|0\right\rangle,\left|\mR\right\rangle)\geq\mC(\left|0\right\rangle,\left|\mR\right\rangle)
\end{align}
for \textit{any} $g_{\pm}$. So any $U(g_\pm)$ (with small $\ep$) applied on the groundstate $\left|0\right\rangle$ (described by the \Poincare-metric) will only increase the complexity with respect to the reference state. We hence see that among the geometries \eqref{metric} in a neighbourhood around the groundstate, the groundstate is the least complex with respect to the reference state $\left|\mR\right\rangle$.
This result has an interesting physical interpretation: The operators $U$ only map states with smooth dual geometry to other states with smooth dual geometry. However, it is commonly assumed that the reference state $\left|\mR\right\rangle$ will be a state without spatial entanglement, and such states cannot have a smooth dual geometry \cite{Bao:2017thr}. So in the total space of states, the states of the form $U(g_+)U(g_-)\left|0\right\rangle$ in a neighbourhood of $\left|0\right\rangle$ form a set in which $\left|0\right\rangle$ locally minimises the complexity with respect to the reference state. See figure \ref{fig::ComplexityOfStates} for an illustrative sketch of the space of states.

The way in which complexity is defined as a distance measure between states can be very abstract and does not need to employ a Riemannian metric, for example it may also be based on Finslerian geometry \cite{2005quant.ph..2070N,2006Sci...311.1133N}. In fact, although one would commonly assume that for a distance measure $\mD$ between states $\phi$ and $\psi$ the symmetry property $\mD(\phi,\psi)=\mD(\psi,\phi)$ has to be satisfied \cite{Susskind:2014jwa}, it was suggested in \cite{Yang:2017nfn} this requirement might have to be abandoned for definitions of complexity. E.g.~one could imagine defining the complexity with respect to a gate-set that includes a given gate, but not its inverse. Then, operators will in general not have the same complexity as their inverse operators.
Furthermore, we consider the distances $\mD(U\left|0\right\rangle,\left|\mR\right\rangle)$ and $\mD(U^\dagger\left|0\right\rangle,\left|\mR\right\rangle)$ between states, which differ in general from each other even if the complexities of the operators $U$ and $U^\dagger$ agree.
However, from the result \eqref{mVresult} it is evident that $\mV_{(2)}$ is invariant under $g_\pm\rightarrow -g_\pm$, which to first order in $\ep$ corresponds to $U(g_\pm)\rightarrow U^\dagger(g_\pm)$, see appendix \ref{sec::AppCFT2}. So, assuming the volume proposal, when applied to the vacuum state $\left|0\right\rangle$ the two operators $U(g_\pm)$ and $U^\dagger(g_\pm)$ lead (at least to leading order in $\ep$) to a change of complexity by the same amount. This may not be true when these operators are applied to generic states or when higher orders of $\ep$ are taken into account.

One of the few kinds of geometric intuition that we can likely rely on when dealing with complexities is the \textit{triangle inequality} \cite{2005quant.ph..2070N,2006Sci...311.1133N}
\begin{align}
\mC(U(g_+)U(g_-)\left|0\right\rangle,\left|\mR\right\rangle)\leq\mC(U(g_+)U(g_-)\left|0\right\rangle,\left|0\right\rangle)+ \mC(\left|0\right\rangle,\left|\mR\right\rangle),
\label{triangle}
\end{align}
hence
\begin{align}
\mC(U(g_+)U(g_-))\geq\mC(U(g_+)U(g_-)\left|0\right\rangle,\left|0\right\rangle)\geq
\mC(U(g_+)U(g_-)\left|0\right\rangle,\left|\mR\right\rangle)-\mC(\left|0\right\rangle,\left|\mR\right\rangle)
\propto \frac{\ep^2\mV_{(2)}}{LG_N}.
\label{triangle2}
\end{align}
So if the proportionality factor in equation \eqref{complexity} could be fixed, our results would lead to quantitative lower bounds on the complexities of the field theory operators $U$. However, as we see from \eqref{mVseries}, the righthand-side of the above bound will be of order $\ep^2$, while for small $\ep$ the change of the state due to the action of $U(g_\pm)$ will be of order $\ep$, and hence we would intuitively assume that $\mC(U(g_+)U(g_-))$ and $\mC(U(g_+)U(g_-)\left|0\right\rangle,\left|0\right\rangle)$ will be of order $\ep$ also. If this is true, the bound of \eqref{triangle2} is not very strict.
It would be interesting to extend our studies to operators $U$ being applied to general Ba\~{n}ados geometries without the need of taking $\ep$ small, however we leave this for the future.

The result \eqref{mVresultb} can be rewritten suggestively as two time-independent pieces, which only depend on $\hat g_+$ or $\hat g_-$, and a time-dependent mixed term
\begin{subequations}
	\begin{align}
	\mV_{(2)} &= \mV_{(2),\text{pure}}(\hat{g}_+) + \mV_{(2),\text{pure}}(\hat{g}_-) + \mV_{(2),\text{mixed}}(\hat{g}_+,\hat{g}_-),
	\label{pureandmixed}
	\\
	\mV_{(2),\text{pure}}(\hat{g}) &= \pi ^3\int\limits_{-\infty}^\infty d\xi ~ |\xi| ^3  \cdot \hat g(-\xi) \hat g(\xi), \\
	\mV_{(2),\text{mixed}}(\hat{g}_+,\hat{g}_-) &= 2 \pi ^3\int\limits_{-\infty}^\infty d\xi ~ |\xi| ^3  \cdot \hat g_+(\xi)\hat g_-(\xi) e^{-4 i \pi \xi t_0}.
	\label{mixed}
	\end{align}
		\label{pureandmixedall}%
\end{subequations}
The reason why this is interesting is that the energy-momentum tensor of the two-dimensional CFT can be understood in terms of left- and right-moving modes. Setting either $g_+=0$ or $g_-=0$ will hence result in a configuration with translational invariance in one of the light-cone coordinates $\tilde{x}^\pm$.
As we integrate over the spatial direction $\tilde{x}$ to obtain the holographic complexity, the result for the change in complexity will then be time-independent and given by either $\mV_{(2),\text{pure}}(\hat{g}_+) $ or $\mV_{(2),\text{pure}}(\hat{g}_-) $ in \eqref{pureandmixed}. If both $g_\pm\neq0$, we get a time-dependent result.

The result \eqref{pureandmixedall} allows for some general insights into the behaviour of $\mV_{(2)}$ under simple manipulations of the functions $\hat{g}_\pm$. For example, we find that under rescalings of arguments, $g_\pm(x) \rightarrow g_\pm(\lambda x)\Leftrightarrow   \hat g_\pm (\xi) \rightarrow \frac{1}{|\lambda|} \hat g_\pm(\xi/\lambda)$,
\begin{subequations}
\begin{align}
  \mV_{\text{pure}} \rightarrow&  \lambda^2 \mV_{\text{pure}} \\
  \mV_{\text{mixed}} \rightarrow&   \lambda^2  \mV_{\text{mixed}}|_{t_0 \rightarrow \lambda t_0}.
\end{align}\label{scalings}%
\end{subequations}

\Comment[inline,caption={Bdy-term}]{
	\begin{minipage}{0.85\linewidth}
under coordinate-shifts 
\begin{align}
	g_\pm(x^\pm) &\rightarrow g_\pm(x^\pm+\delta_\pm)  \Rightarrow
	\hat g_\pm(\xi) \rightarrow \hat g_\pm(\xi) e^{-2\pi i \xi \delta_\pm},
\end{align}
we find
\begin{subequations}
	\begin{align}
	V_{(2),\text{pure}}(\hat g(\xi) e^{-2\pi i \xi \delta}) &= V_{(2),\text{pure}}(\hat g(\xi))\\
	V_{(2),\text{mixed}}(\hat g_+(\xi) e^{-2\pi i \xi \delta_+},\hat g_+(\xi) e^{-2\pi i \xi \delta_-}) &= \left.V_{(2),
\text{mixed}}(\hat g_+(\xi),\hat g_-(\xi))\right|_{t_0 \rightarrow t_0 + (\delta_+ + \delta_-)/2} \\[0.5em]
	\Rightarrow	V_{(2)}(\hat g_+(\xi) e^{-2\pi i \xi \delta_+},\hat g_+(\xi) e^{-2\pi i \xi \delta_-}) &= \left.V_{(2)}(\hat g_+(\xi),\hat g_-(\xi))\right|_{t_0 \rightarrow t_0 + (\delta_+ + \delta_-)/2}.
	\end{align}
\end{subequations}

special cases
\begin{subequations}
\begin{align}
 &\text{time-shift:} &   & \text{translation: }\nonumber \\[0.5em]
  g_\pm(x^\pm) &\rightarrow g_\pm(x^\pm+\delta) & g_\pm(x^\pm) &\rightarrow g_\pm(x^\pm\pm \delta) \\
  \hat g_\pm(\xi) &\rightarrow \hat g_\pm(\xi) e^{-2\pi i \xi \delta} & \hat g_\pm(\xi) &\rightarrow \hat g_\pm(\xi) e^{\mp 2\pi i \xi \delta} \\
  V_{(2)} &\rightarrow \left. V_{(2)}\right|_{t_0 \rightarrow t_0 + \delta } & V_{(2)} &\rightarrow  V_{(2)}
\end{align}
\end{subequations}
\end{minipage}
}

Another interesting thing to look at is what happens under addition of functions $g_\pm$, as at first order in $\ep$ this corresponds to carrying out two small conformal transformations after each other (see appendix \ref{sec::products}).  We obtain
\begin{subequations}
	\begin{align}
	  \mV_{(2),\text{pure}}(\hat g + \hat f) &=\mV_{(2),\text{pure}}(\hat g)+ \mV_{(2),\text{pure}}(\hat f) + 2\pi^3 \int d\xi |\xi|^3 \hat g(-\xi) \hat f(\xi)\\
	  \mV_{(2),\text{mixed}}(\hat g_+ + \hat f_+,\hat g_- + \hat f_-) &= \mV_{(2),\text{mixed}}(\hat g_+ ,\hat g_- )+ \mV_{(2),\text{mixed}}( \hat f_+, \hat f_-)\nonumber \\
	  &~~~+ \mV_{(2),\text{mixed}}( \hat f_+,\hat g_-) + \mV_{(2),\text{mixed}}(\hat g_+,\hat f_-).
	\end{align}\label{addition}%
\end{subequations}

\Comment[inline,caption={Further property}]{
	\begin{minipage}{0.85\linewidth}
			\begin{align}
			V_{(2),\text{pure}}(\hat g(\xi)  + \hat f(\xi) e^{-2\pi i \xi \delta}) &= V_{(2),\text{pure}}(\hat g)+ V_{(2),\text{pure}}(\hat
		f) + 2\pi^3 \int d\xi |\xi|^3 \hat g(-\xi) \hat f(\xi)e^{-2\pi i \xi \delta} \nonumber\\
			&= \left. V_{(2)}(\hat g(-\xi),\hat f(\xi))\right|_{t_0 \rightarrow \delta/2}
			\end{align}

		\begin{align}
		&V_{(2),\text{mixed}}\qty(\hat g_+(\xi)  + \hat f_+(\xi) e^{-2\pi i \xi \delta_+},\hat g_-(\xi)  + \hat f_-(\xi) e^{-2\pi i \xi \delta_-}) \nonumber\\ &=V_{(2),\text{mixed}}(\hat g_+ ,\hat g_- )+ \left. V_{(2),\text{mixed}}( \hat f_+, \hat f_-)\right|_{t_0 \rightarrow t_0 + (\delta_+ + \delta_-)/2}\nonumber\\
		&~~~+ \left.V_{(2),\text{mixed}}( \hat f_+,\hat g_-) \right|_{t_0 \rightarrow t_0 + \delta_+/2} + \left.V_{(2),
\text{mixed}}(\hat g_+,\hat f_-)\right|_{t_0 \rightarrow t_0 +  \delta_-/2}
		\end{align}\
\end{minipage}
}
In the next section, we will proceed to pick some illustrative examples of $g_\pm$ which allow for particularly simple analytical expressions to be derived for $t_{1}(\tilde{x},\tilde{\lambda})$ and $\mV_{(2)}$.

%% file: ComplexityOfStates2.pdf_tex
\begingroup%
  \makeatletter%
  \providecommand\color[2][]{%
    \errmessage{(Inkscape) Color is used for the text in Inkscape, but the package 'color.sty' is not loaded}%
    \renewcommand\color[2][]{}%
  }%
  \providecommand\transparent[1]{%
    \errmessage{(Inkscape) Transparency is used (non-zero) for the text in Inkscape, but the package 'transparent.sty' is not loaded}%
    \renewcommand\transparent[1]{}%
  }%
  \providecommand\rotatebox[2]{#2}%
  \ifx\svgwidth\undefined%
    \setlength{\unitlength}{695.10103175bp}%
    \ifx\svgscale\undefined%
      \relax%
    \else%
      \setlength{\unitlength}{\unitlength * \real{\svgscale}}%
    \fi%
  \else%
    \setlength{\unitlength}{\svgwidth}%
  \fi%
  \global\let\svgwidth\undefined%
  \global\let\svgscale\undefined%
  \makeatother%
  \begin{picture}(1,0.56614217)%
    \put(0,0){\includegraphics[width=\unitlength,page=1]{ComplexityOfStates2.pdf}}%
    \put(0.97112776,0.86869857){\color[rgb]{0,0,0}\makebox(0,0)[lt]{\begin{minipage}{0.77482647\unitlength}\raggedright \end{minipage}}}%
    \put(0.32995504,0.54855255){\color[rgb]{0,0,0}\makebox(0,0)[lb]{\smash{$\left|\psi_{U}\right\rangle$}}}%
    \put(0.01057153,0.00542862){\color[rgb]{0,0,0}\makebox(0,0)[lb]{\smash{$\left|\mathcal{R}\right\rangle$}}}%
    \put(0.00578456,0.54733666){\color[rgb]{0,0,0}\makebox(0,0)[lb]{\smash{$\left|0\right\rangle$}}}%
    \put(0.13861338,0.48911403){\color[rgb]{0,0,0}\makebox(0,0)[lb]{\smash{$U$}}}%
    \put(0.16395922,0.23278113){\color[rgb]{1,0,0}\makebox(0,0)[lb]{\smash{$\mathcal{C}(\left|\psi_{U}\right\rangle,\left|\mathcal{R}\right\rangle)$}}}%
    \put(0.01459994,0.40810449){\color[rgb]{1,0,0}\makebox(0,0)[lb]{\smash{$\mathcal{C}(\left|0\right\rangle,\left|\mathcal{R}\right\rangle)$}}}%
  \end{picture}%
\endgroup%

%% file: NinasResults.tex
\section{Examples}
\label{sec::Examples}

\subsection*{Example 1}
\begin{figure}[htb]
	\centering

	\includegraphics[width=0.5\linewidth]{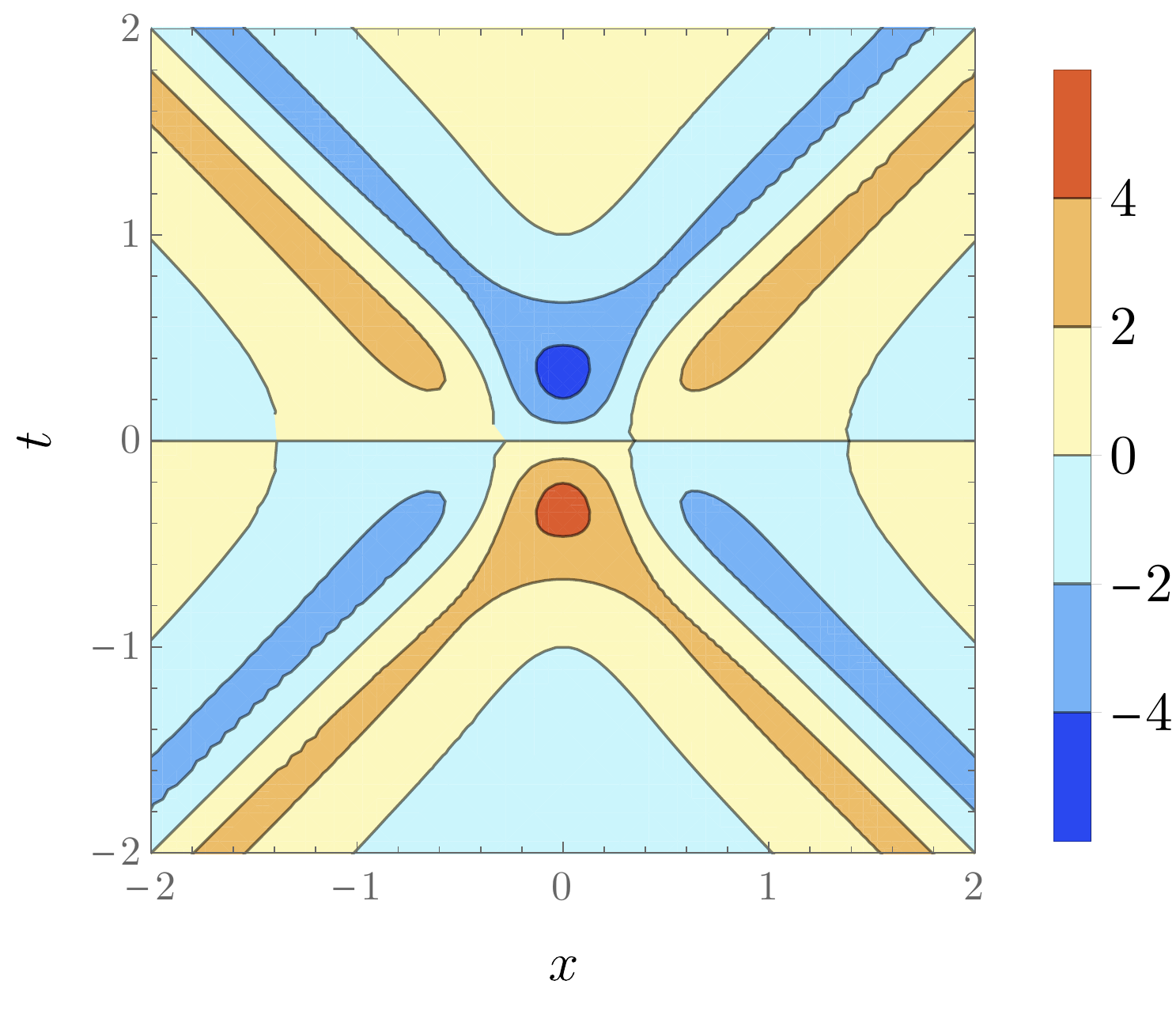}
	\caption{First order term in $\ep$ (as in \eqref{Ttosecondorder}) of the energy density $\mE(t,x)=T_{++}(t+x) + T_{--}(t-x)$ for example 1 with $a_\pm=c_\pm=1$. This picture can be interpreted in terms of two wavepackets, one moving to the left and one moving to the right at the speed of light.
	}
	\label{fig:EnergyDensity}
\end{figure}

To begin, let us choose\footnote{For simplicity of notation, we will drop the tildes over the coordinates in this section.}
\begin{subequations}
\begin{align}
	g_{+}\left(x^+\right)= \frac{a_+ c_+}{a_+^2+\left(x^+\right)^2} & \Leftrightarrow \hat{g}_{+}\text{($\xi $)=}  c_+ e^{-2 a_+ \pi  \left| \xi \right| } \pi,  \\
	g_{-}\left(x^-\right)=\frac{a_- c_-}{a_-^2+\left(x^-\right)^2} & \Leftrightarrow \hat{g}_{-}\text{($\xi $)=}  c_- e^{-2 a_- \pi  \left| \xi \right| } \pi,
\end{align}
\end{subequations}
where $a_\pm$ are positive.
These functions have the virtue of allowing for comparably simple analytic expressions. Furthermore, they fall off towards infinity like a power law and hence describe wavepackets which are approximately localised, see figure \ref{fig:EnergyDensity}.

For the first order correction to the embedding in \eqref{tseries}, we find
\begin{align}
t_1( x,\lambda)  =& c_-\frac{ 2 \sqrt{\lambda } \left((t_0-x)^2-a_-^2\right)-\lambda a_- \left(a_-^2-3(t_0-x)^2\right)-a_-}{2 \left(a_-^2+(t_0-x)^2\right) \left(2 a_- \sqrt{\lambda }+\lambda  \left(a_-^2+(t_0-x)^2\right)+1\right){}^2} \nonumber \\
& + c_+ \frac{2 \sqrt{\lambda } \left((t_0+x)^2-a_+^2\right)-\lambda a_+ \left(a_+^2-3(t_0+x)^2\right)-a_+}{2 \left(a_+^2+(t_0+x)^2\right) \left(2 a_+ \sqrt{\lambda }+\lambda  \left(a_+^2+(t_0+x)^2   \right)+1\right){}^2}
\end{align}
and the correction to the volume is given by
\begin{align}
\mV_{(2)} =& c_+^2\cdot \frac{3 \pi }{64 a_+^4} + c_-^2\cdot \frac{3 \pi }{64 a_-^4} + \frac{c_- c_+}{(a_+ +a_-)^4} \cdot \frac{3 \pi  \left(16 t_a^4-24 t_a^2+1\right)}{2 \left(4 t_a^2+1\right){}^4} \ \  \text{with } t_a = \frac{t_0}{a_+ + a_-},
\label{Ex1mV}
\end{align}
which nicely displays the structure derived in \eqref{pureandmixed}. The mixed term $\mV_{(2),\text{mixed}}$ is shown in figure \ref{fig:Example1}.

\subsection*{Example 2}
\begin{figure}\centering
	\includegraphics[width=0.65\linewidth]{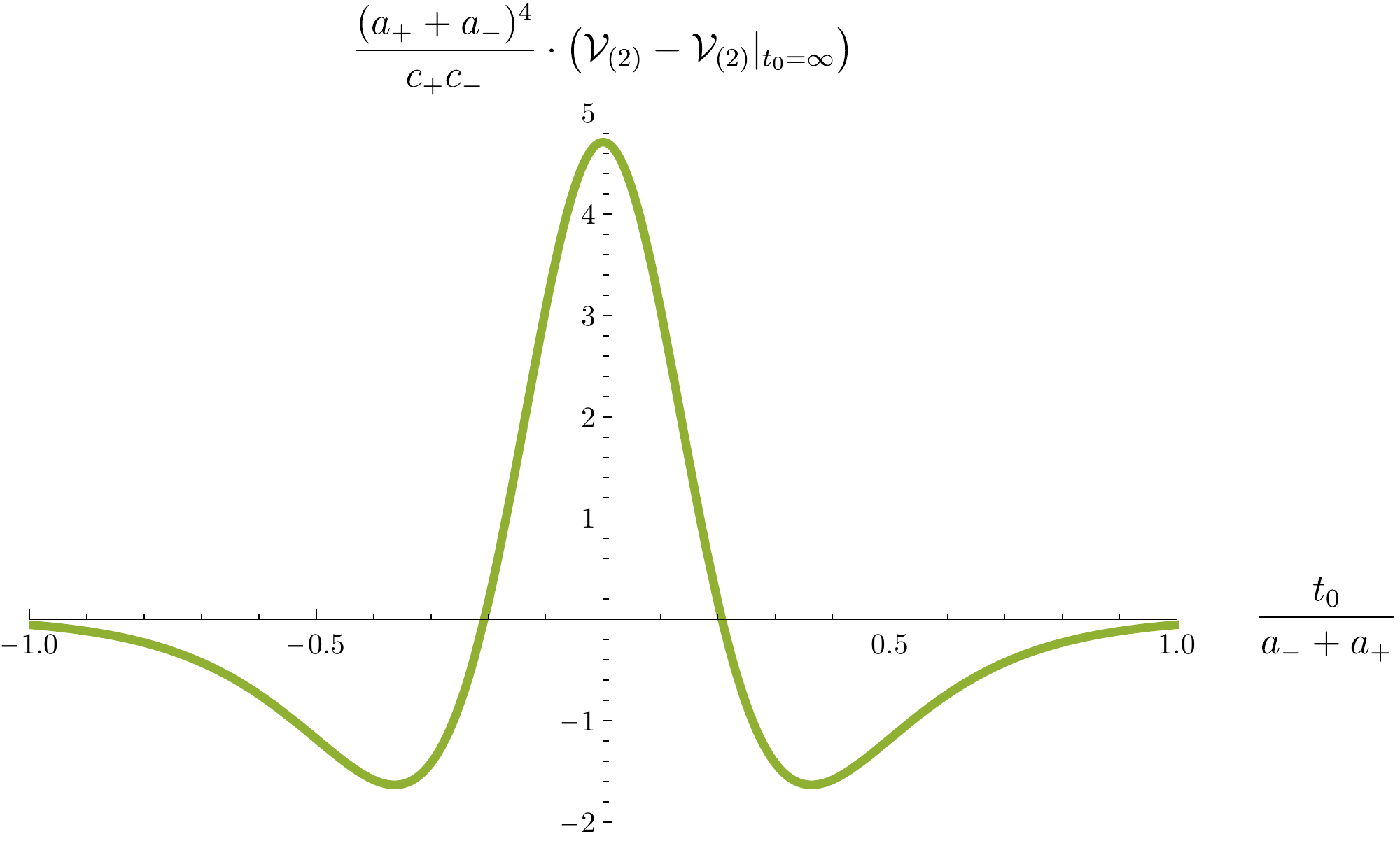}
	\caption{Plot of the term $\mV_{(2),\text{mixed}}(\hat{g}_+,\hat{g}_-)$ for example 1, respectively $-\mV_{(2),\text{mixed}}(\hat{g}_+,\hat{g}_-)$ for example 2. For $t_0\rightarrow\pm\infty$, $\mV_{(2),\text{mixed}}(\hat{g}_+,\hat{g}_-)\rightarrow0$.
	}
	\label{fig:Example1}
\end{figure}

Using instead
\begin{subequations}
\begin{align}
	g_{+}\left(x^+\right)=  -\frac{c_+ x^+}{a_+^2+\left(x^+\right)^2} &\Leftrightarrow \hat{g}_{+}\text{($\xi $)=} -i c_+ e^{-2 a_+ \pi  \left| \xi \right| } \pi  \text{sgn}(\xi ) \\
	g_{-}\left(x^-\right)=  -\frac{c_- x^-}{a_-^2+\left(x^-\right)^2} &\Leftrightarrow \hat{g}_{-}\text{($\xi $)=} -i c_- e^{-2 a_- \pi  \left| \xi \right| } \pi  \text{sgn}(\xi ),
\end{align}
\end{subequations}
we obtain the embedding
\begin{align}
t_1( x,\lambda)   =&- c_- \frac{ \left(t_0-x\right) \left(-4 a_- \sqrt{\lambda }+\lambda  \left(-3 a_-^2+(t_0-x)^2\right)-1\right)}{2 \left(a_-^2+(t_0-x)^2\right) \left(2 a_- \sqrt{\lambda }+\lambda  \left(a_-^2+(t_0-x)^2\right)+1\right){}^2} \nonumber\\
&-c_+ \frac{\left(t_0+x\right) \left(-4 a_+ \sqrt{\lambda }+\lambda  \left(-3 a_+^2+(t_0+x)^2\right)-1\right)}{2 \left(a_+^2+(t_0+x)^2\right) \left(2 a_+ \sqrt{\lambda }+\lambda  \left(a_+^2+(t_0+x)^2   \right)+1\right){}^2}
\end{align}
and the volume change
\begin{align}
\mV_{(2)} =& c_+^2\cdot \frac{3 \pi }{64 a_+^4} + c_-^2\cdot \frac{3 \pi }{64 a_-^4} - \frac{c_- c_+}{(a_+ +a_-)^4} \cdot \frac{3 \pi  \left(16 t_a^4-24 t_a^2+1\right)}{2 \left(4 t_a^2+1\right){}^4} \ \  \text{with } t_a = \frac{t_0}{a_+ + a_-}.
\label{Ex2mV}
\end{align}
Interestingly, this is identical to the result \eqref{Ex1mV} up to a sign change in the term $\mV_{(2),\text{mixed}}(\hat{g}_+,\hat{g}_-)$. See figure \ref{fig:Example1} for a plot of this quantity.

Interestingly, we see that $\mV_{(2),\text{mixed}}(\hat{g}_+,\hat{g}_-)\rightarrow0$ as $t_0\rightarrow\pm\infty$. The physical reason for this appears to be that the wavepackages of left and right moving modes propagate away from each other in this limit and have increasingly little overlap due to the falloff of the functions $g_\pm(x^\pm)$ on a constant time slice. The larger the separation of these wavepackages, the less they influence each other, and in the limit $t_0\rightarrow\pm\infty$ the change in complexity $\mV_{(2)}$ becomes the sum of the complexity changes due to each individual wavepackage:
\begin{align}
\lim_{t_0\rightarrow\pm\infty}\mV_{(2)} \rightarrow \mV_{(2),\text{pure}}(\hat{g}_+) + \mV_{(2),\text{pure}}(\hat{g}_-).
\label{limit}
\end{align}
Another noteworthy feature of the time-dependent contribution is that $\int\limits_{-\infty}^{+\infty}\mV_{(2),\text{mixed}}(\hat{g}_+,\hat{g}_-) dt_0=0$, which is a generic consequence of \eqref{mixed} for sufficiently well-behaved $\hat{g}_{\pm}$. This means that the time-average of $\mV_{(2)} $ is also given by the sum on the right-hand side of \eqref{limit}.

These additivity properties are interesting in the light of the axiom \textbf{G3} proposed in \cite{Yang:2018nda}, which qualitatively states that the sum of the complexity of two independent tasks should be the sum of the complexities of each individual task. For non-zero functions $g_\pm$, the conformal transformation on the CFT state is implemented by the two commuting operators $U(g_+)\equiv U_L$ and $U(g_-)\equiv U_R$, as there are \textit{two} copies of the Virasoro algebra.
By \textbf{G3} of \cite{Yang:2018nda}, we would hence expect
\begin{align}
\mC(U_LU_R)=\mC(U_L)+\mC(U_R).
\end{align}
However, due to the triangle inequality \eqref{triangle} and the inequality explained in footnote \ref{f1}, this does not translate directly into a statement about the complexities of the states $U_LU_R\left|0\right\rangle$, so the fact that the additivity of \eqref{limit} does not hold for finite times is not inconsistent.

\subsection*{Example 3}
\begin{figure}\centering
	\includegraphics[width=0.65\linewidth]{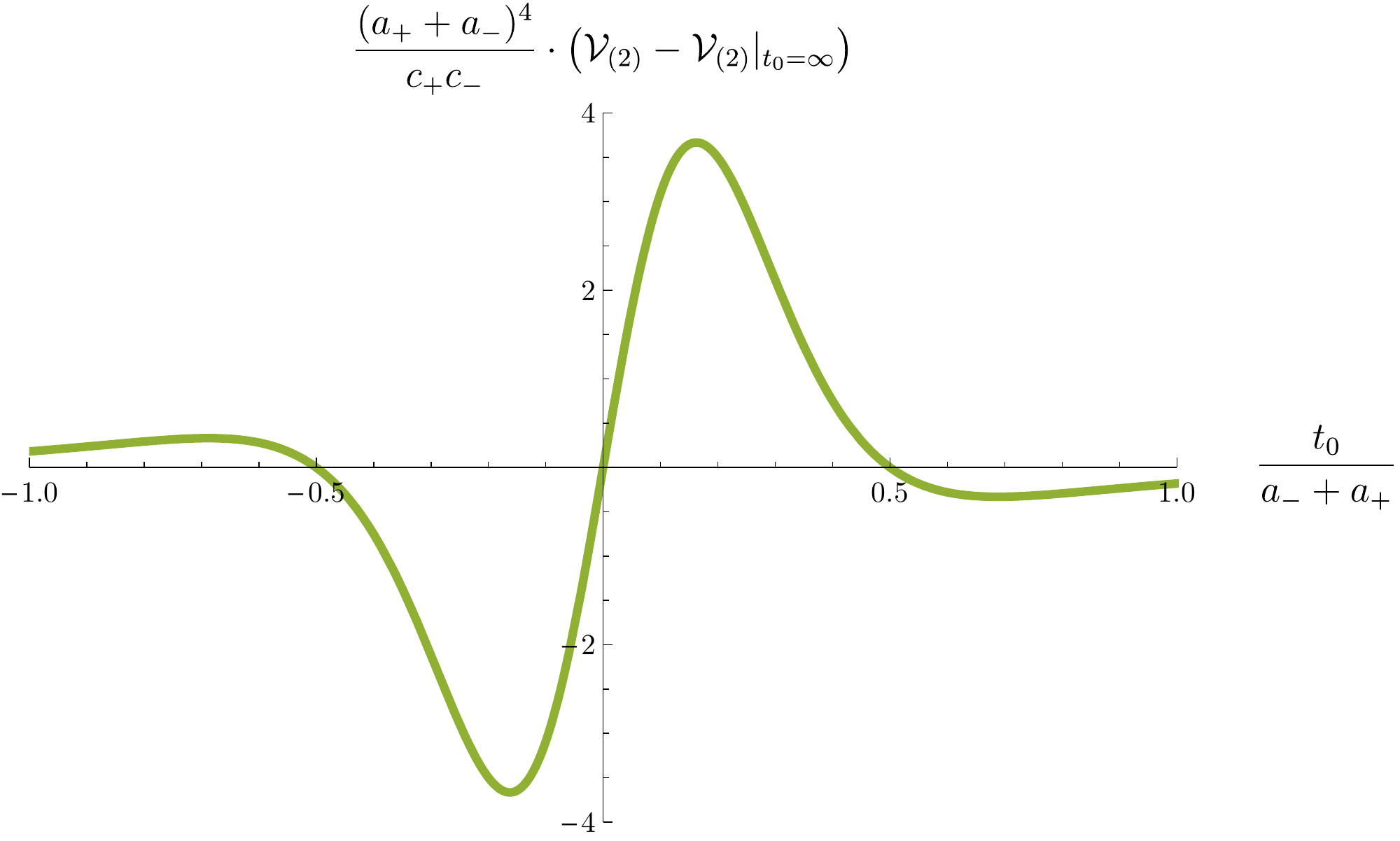}
	\caption{$\mV_{(2),\text{mixed}}(\hat{g}_+,\hat{g}_-)$ for example 3.}
	\label{fig:Example3}
\end{figure}

As a final example, we may look at
\begin{subequations}
\begin{align}
	g_{+}\left(x^+\right)= \frac{c_+ x^+}{a_+^2+\left(x^+\right)^2} &\Leftrightarrow \hat{g}_{+}\text{($\xi $)=}  i c_+ e^{-2 a_+ \pi |\xi| } \pi\text{sgn}(\xi)   \\
	g_{-}\left(x^-\right)= \frac{a_- c_-}{a_-^2+\left(x^-\right)^2} & \Leftrightarrow \hat{g}_{-}\text{($\xi $)=}  c_- e^{-2 a_- \pi  |\xi| } \pi ,
\end{align}
\end{subequations}
which yields
\begin{align}
\mV_{(2)} =& c_+^2\cdot \frac{3 \pi }{64 a_+^4} + c_-^2\cdot \frac{3 \pi }{64 a_-^4} - \frac{c_- c_+}{(a_+ +a_-)^4} \cdot \frac{12 \pi  t_a \left(1-4 t_a^2\right)}{\left(4 t_a^2+1\right){}^4} \ \  \text{with } t_a = \frac{t_0}{a_+ + a_-}
\end{align}
The time-dependent part $\mV_{(2),\text{mixed}}(\hat{g}_+,\hat{g}_-)$ of this expression is plotted in figure \ref{fig:Example3}. As before, this quantity vanishes in the limit $t_0\rightarrow\pm\infty$.

%% file: axiomatic.tex
\section{Comparison to field theory ansatz}
\label{sec::FTproposal}

\renewcommand{\ket}{\right\rangle}
\renewcommand{\bra}{\left\langle}

In this section, we will compare our results to a field-theory complexity proposal made in ArXiv version 1 of \cite{Yang:2018nda} and \cite{YangNext}. There, it was proposed that \textit{axiomatic complexity} between two pure states $\left| \psi_1\ket$ and $\left| \psi_2\ket$ could be defined via their fidelity as
\begin{align}
\mC\left(\psi_1,\psi_2\right)=-2\ln(|\bra\psi_1|\psi_2\ket |).
\label{Yang}
\end{align}
Irrespectively of whether this proposal is the correct result for the field theory dual of \textit{holographic complexity} (see the discussion in \cite{Yang:2018nda,YangNext} and, however, \cite{Brown:2017jil,Susskind:2018pmk} for arguments as to why this would not be a good measure of complexity in the sense of \textit{circuit complexity}), we will use \eqref{Yang} as an easy to work with toy-model for field theory complexity, and we will derive the non-trivial constraints on the reference state $\left| \mR \ket$ that our results of section \ref{sec::CV} imply when taken together with \eqref{Yang}.

First of all, for small $\ep$, we expand the operator generating a conformal transformation as
\begin{align}
U(\ep)\approx 1+\sigma U_1 + \sigma^2 U_2+...
\end{align}
with\footnote{Where, from appendix \ref{sec::AppCFT2}, specifically section \ref{sec::realspace}, we take
	\begin{align}
	\mL &= \ep ~\frac{i}{2\pi} \int dx~ g_+(x)\cdot T_{++}(x),
\nonumber
\\
	\mB &=-\ep^2 ~\frac{i}{8\pi} \iint dx_1 dx_2~ g_+'(x_1)g_+(x_2)\cdot T_{++}(x_1+x_2).
\nonumber
	\end{align}
	For these, we have $\mL^\dagger = -\mL,\  \mB^\dagger =-\mB$.},
\begin{align}
\ep U_1&=\mL,
\ \ 
\ep^2 U_2
=\frac{1}{2}\mL^2+\mB.
\end{align}
We are implicitly still making the assumption that the functions $g_\pm(x_\pm)$ on which these operators will depend are smooth and well behaved enough to allow for all the manipulations performed in the earlier sections of the paper, i.e. for all integrals to converge etc. The operators $U_1, U_2, \mL$ and $\mB$ hence stand for a large family of operators which are dependent on a choice of functions $g_\pm(x_\pm)$ which is arbitrary to a degree within these limitations.
We can now expand\footnote{Here we explicitly assume that the reference state $\left|\mR\ket$ is a pure state. This might not be necessarily true, we thank Keun-Young Kim for discussions on this issue, see also  \cite{Yang:2018nda}. In the main results of this section, such as \eqref{condition} for example, we can however re-generalise to a mixed reference state by making the replacement $\left|\mR\ket\bra\mR\right|\rightarrow \rho_\mR$.}
\begin{align}
\bra \mR| U(\ep)|0\ket \approx \bra \mR|0\ket+\ep \bra \mR|U_1|0\ket+\ep^2 \bra \mR|U_2|0\ket....
\end{align}
For the fidelity, this then yields
\begin{align}
|\bra \mR|U(\ep)|0\ket |^2 &= \bra \mR|U(\ep)|0\ket \bra 0 |U^\dagger(\ep)|\mR\ket \nonumber
\\
&=A_0+ \ep A_1
+ \ep^2 A_2
\end{align}
with
\begin{align}
A_0&\equiv\bra \mR|0\ket \bra 0|\mR\ket,
\label{A_0}
\\
A_1&\equiv\bra \mR|0\ket\bra 0|U_1^\dagger|\mR\ket+\bra 0|\mR\ket \bra \mR|U_1|0\ket,
\\
A_2&\equiv\bra \mR|0\ket\bra 0|U_2^\dagger|\mR\ket+\bra 0|\mR\ket \bra \mR|U_2|0\ket+\bra 0|U_1^\dagger|\mR\ket \bra \mR|U_1|0\ket.
\label{A_2}
\end{align}

Plugging this into \eqref{Yang}, we find
\begin{align}
\mC\left(U(\ep)\left|0\ket,\left|\mR\ket\right)&=-\log\left(|\bra\mR|U(\ep)|0\ket|^2 \right)\nonumber
\\
&\approx-\log\left(A_0+\ep A_1+\ep^2 A_2\right)\nonumber
\\
&\approx -\log(A_0) -\ep\frac{A_1}{A_0} + \ep^2 \qty(\frac{1}{2} \qty(\frac{A_1}{A_0})^2-\frac{A_2}{A_0}) \label{Bexp}
\end{align}.

We are now in a position to compare \eqref{Bexp} to our results obtained in the earlier sections assuming the holographic volume proposal \eqref{complexity}. Most importantly, \eqref{mVseries} seems to show that the first order term $-\ep \frac{2B_1}{B_0}$ should vanish in \eqref{Bexp}. However, if \eqref{Yang} is to hold in the case of a continuum field theory, we have to be careful about the appearance of the UV cutoff $\epsilon \ll 1$. Holography shows \cite{Carmi:2016wjl}
\begin{align}
\mC\left(\left|0\ket,\left|\mR\ket\right)\sim \frac{1}{\epsilon}\gg 1.
\end{align}
We argued that complex phases should not affect the value of complexity, and this is certainly true in the proposal \eqref{Yang}. So without loss of generality, we can choose the complex phase of $\left|\mR\ket$ such that
\begin{align}
|\bra\mR|0\ket| =\bra\mR|0\ket= \bra 0|\mR\ket = e^{-\frac{1}{2}\mC\left(\left|0\ket,\left|\mR\ket\right)}\sim e^{\frac{-1}{2\epsilon}}.
\end{align}
Due to the exponentiation in the last step, it is conceivable that some of the terms in \eqref{Bexp} might not vanish, but still be nonperturbatively small, and might hence not be captured by a holographic calculation as done in the previous sections. However, the vanishing of the first order term in \eqref{Bexp} is not only implied by \eqref{mVseries}, but also by the general expectation that the complexity change due to $U$ or the inverse $U^\dagger$ being applied to the ground state $\left|0\ket$ should be identical, at least up to second order (see section \ref{sec::CV} and also \cite{Yang:2018nda}). We hence obtain
\begin{align}
\frac{A_1}{A_0} \equiv 0 \Leftrightarrow \Re\left(\frac{\bra 0| \mR\ket\bra \mR|U_1|0\ket }{\bra0|\mR\ket\bra \mR | 0 \ket}\right)
\equiv0
\label{condition}
\end{align}%
This is the first and most important condition on the reference state $\left|\mR\right>$ that we derive in this section. It should hold for any sufficiently well behaved choice of $g_\pm(x_\pm)$, and hence represents an entire class of restrictions on the reference state.

We now turn our attention to the second order term in \eqref{Bexp}:
\begin{align}
\frac{A_1^2}{2A_0^2}-\frac{A_2}{A_0}
\end{align}
Assuming \eqref{condition}, this simplifies to
\begin{align}
-\frac{A_2}{A_0}
&=-{2}\Re\left( \frac{ \bra 0 | \mR \ket \bra \mR|U_2|0\ket}{\bra0|\mR\ket \bra \mR|0\ket}\right)+\frac{\bra 0|U_1^\dagger|\mR\ket \bra \mR|U_1|0\ket}{\bra0|\mR\ket \bra \mR|0\ket} \nonumber
\\
&=\frac{-2 \Re\left(  \bra0|\mR\ket \bra \mR|U_2|0\ket \right)+| \bra \mR|U_1|0\ket |^2}{\bra0|\mR\ket \bra \mR|0\ket}
\label{SecondOrder}
\end{align}

As argued above, we expect the complexity change induced by acting on the groundstate with
\begin{align}
U&\approx 1+\underbrace{\mL}_{\equiv \ep U_1} +\underbrace{\mB+\frac{1}{2}\mL^2}_{\equiv\ep^2 U_2}+...
\end{align}
to be identical to the complexity change induced by acting on the groundstate with
\begin{align}
U^\dagger&\approx 1\underbrace{-\mL}_{\equiv\ep U_1} \underbrace{-\mB+\frac{1}{2}\mL^2}_{\equiv\ep^2 U_2}+...\ .
\end{align}
This implies
\begin{align}
&{2}\Re\left(\frac{\bra 0|\mR\ket \bra \mR|\mB+\frac{1}{2}\mL^2|0\ket}{\bra0|\mR\ket\bra \mR | 0 \ket}\right)+\frac{|\bra \mR|\mL|0\ket|^2}{\bra0|\mR\ket\bra \mR | 0 \ket}
\equiv
{2}\Re\left(\frac{\bra 0|\mR\ket \bra \mR|-\mB+\frac{1}{2}\mL^2|0\ket}{\bra0|\mR\ket\bra \mR | 0 \ket}\right)+\frac{|\bra \mR|-\mL|0\ket|^2}{\bra0|\mR\ket\bra \mR | 0 \ket}
\label{equality}
\\
&\Leftrightarrow \Re\qty(\frac{\bra 0|\mR\ket\bra \mR|\mB|0\ket}{\bra0|\mR\ket\bra \mR | 0 \ket}) \equiv0
\label{cond2}
\end{align}

As the condition \eqref{condition} is supposed to hold for any reasonable choice of function $g_\pm$, and as otherwise the two operators $\mL$ and $\mB$ have a similar structure (see appendix \ref{sec::AppCFT2}), \eqref{cond2} actually follows from \eqref{condition}. Reversing the above reasoning, \eqref{equality} is hence implied by \eqref{condition}, too. Consequently, the second order term in the complexity change only depends on the term $\sim \mL^2$ in the expansion of the operator to second order, and not the term $\sim \mB$, as may have been anticipated from the discussion in section \ref{sec::CV}.

Consequently, in \eqref{SecondOrder}, the positivity of our result \eqref{mVresult} and its independence of the UV cutoff imply
\begin{align}
-{2}\Re\left(\bra 0|\mR\ket \bra \mR|\frac{1}{2}\mL^2|0\ket\right)\geq|\bra \mR|\mL|0\ket|^2,
\\
{2}\Re\left(\bra 0|\mR\ket \bra \mR|\frac{1}{2}\mL^2|0\ket\right)+|\bra \mR|\mL|0\ket|^2\sim -e^{-\mC\left(\left|0\ket,\left|\mR\ket\right)}.
\end{align}

These appear to be non-trivial conditions on the reference state $\left|\mR\ket$. We leave it for future research to investigate these conditions and the possible reference states that satisfy them in more detail.


%% file: Conc.tex
\section{Conclusion and outlook}
\label{sec::Conc}

To summarise: Based on the volume proposal \eqref{complexity} for holographic complexity, and the solution generating diffeomorphisms explained in section \ref{sec::SGD} for small $\ep$, we have calculated explicit and general formulas \eqref{mVresult}, \eqref{pureandmixedall} for how holographic complexity changes when such transformations are applied to the ground state in section \ref{sec::CV}.
We have also pointed out a number of nontrivial analytic properties of these expressions, such as the formulas for  scaling \eqref{scalings} and the addition \eqref{addition}.
In section \ref{sec::Examples}, we studied a couple of explicit and analytically solvable examples which exemplify the generic results of section \ref{sec::CV}. Also, in \eqref{limit} we pointed out the interesting property of additivity of both the late time results and the time-average of the change in complexity when using conformal transformations that make use of both copies of the Virasoro group.

The motivation behind our investigations was the idea that if holographic complexity is really a measure of a field theoretic concept of complexity along the lines explained in section \ref{sec::Intro}, then there have to be some \textit{implicit} choices of e.g.~gateset or reference state associated with it. When comparing our results to how complexity changes under small conformal transformations in field theory proposals for complexity, this might help to shed light on the choices that are implicit in the holographic proposal. We started with some preliminary work in this direction in section \ref{sec::FTproposal}.
In this context it is important that, of course, conformal transformations are something that can be studied in \textit{any} CFT, not only strongly coupled or large $c$ ones with a classical gravity dual. This should make a comparison with field theory results easier in general\footnote{Note: After the first version of this paper was released, a proposal was made in \cite{Belin:2018bpg} for a complexity-like field-theory quantity dual to bulk volumes, and in fact that paper found precise agreement with our results.}.

Let us close with an outlook on interesting investigations that may now follow. Firstly, it would obviously be of great interest to study how complexity changes under conformal transformations according to the field theory prescriptions of e.g.~of \cite{Jefferson:2017sdb,Chapman:2017rqy,Caputa:2017urj,Caputa:2017yrh,Hashimoto:2018bmb,Reynolds:2017jfs,Yang:2017nfn} and compare the results to the ones presented in this paper, as well as continue the work on the proposal of \cite{Yang:2018nda,YangNext} started in section \ref{sec::FTproposal}.
Similarly, it would be interesting to compare these results to similar results that could be gained from the holographic action proposal \eqref{action}\footnote{Note: After the first version of this paper was released, the calculation of complexity change under conformal transformations according to the action proposal was completed and published in \cite{Flory:2019kah}. There, it was shown that the behaviour of the action proposal under such transformations is qualitatively different from the volume proposal, and that certain field theory candidates for definitions of complexity cannot be dual to holographic complexity via the action proposal.}.

Secondly, another worthwhile direction may be to study the effect of (infinitesimal) conformal transformations on \textit{subregions} of the dual CFT$_{2}$. When acting with such a small transformation on the groundstate, the resulting energy-momentum tensor \eqref{Ttosecondorder} will necessarily  violate the null energy condition (NEC) \textit{somewhere}.\footnote{Due to the assumption that $g_\pm \rightarrow 0$ asymptotically, its third derivative cannot be non-positive everywhere, so for $\ep\ll1$ \eqref{Ttosecondorder} implies $T_{\pm\pm}<0$ for some $\tilde{x}_\pm$.}
This is not problematic, as in holography the dual field theory is a \textit{quantum} field theory, and it is well known that quantum effects can lead to a violation of the NEC. In fact holography has been used as a tool to study such NEC violations on the boundary in the past. 
Now, out of phyiscal intuition, it might be interesting to ask whether subregions in which the NEC is violated have a larger or lower complexity than similarly sized and shaped ones in which the average energy is positive.

Thirdly, one could holographically calculate the complexities of the states corresponding to the two sided black holes studied in  \cite{Mandal:2014wfa}, i.e.~BTZ black holes to which a solution generating diffeomorphism was applied on one or both of its sides. This way one generates an infinite set of purifications of the thermal state, of which the well known thermofield-double state is only one example. One might then ask: Which of these states is the optimal purification of the thermal state in the sense of having minimal complexity, and what makes this purification special?

We leave these interesting questions, as well as many others, for future research.

%% file: AppendixCFT2New.tex
\section{Unitary operator for conformal transformations}
\label{sec::AppCFT2}

In this appendix we will review how the unitary operators implementing conformal transformations in equation \eqref{Uconf} can be explicitly constructed in terms of field theory operators, specifically in terms of the Virasoro generators $L_n$. 
For this, we will follow the perturbative approach of appendix E of \cite{Mandal:2014wfa} and extend it to second order. 

\subsection{Conformal maps}
\label{sec::maps}

Let us start by considering a compact spatial direction, i.e.~$x \in [0,\ell]$. The usual coordinates for radial quantisation are
\begin{subequations}
	\begin{align}
	w = \exp \left(i\frac{2 \pi}{\ell}\cdot x^+\right), 
	\label{wisexp}
	\\
	\bar w = \exp \left(i\frac{2 \pi}{\ell}\cdot x^-\right).
	\end{align} 
\end{subequations}
In the following, we demand $|w| = 1 \Leftrightarrow w^\dagger = w^{-1}$ to ensure $x^\pm\in\mathbb{R}$. 
The energy-momentum tensor in these radial coordinates is\footnote{Of course, this is only the chiral part of the energy-momentum tensor, similar relations hold for the anti-chiral part $\bar{T}(\bar{w})$.}
\begin{subequations}
	\begin{align}
	T(w) =& \sum_n L_n w^{-n-2} ,
	\end{align}
	and in the light-cone coordinates\footnote{The $const.$ is related to the Casimir energy on the cylinder. Note however that later we will send $\ell\rightarrow\infty$.}
	\begin{align}
	T_{++} = -\frac{4\pi^2}{\ell^2} \left(w^2 T(w)+const. \right). \label{EMTLC}
	\end{align}
\end{subequations}
%
This expression has to be real, yielding the condition $L_n ^\dagger = L_{-n}$ for the \textit{Virasoro generators} $L_n$, which satisfy the well known algebra
\begin{subequations}
	\begin{align}
\left[L_{m},L_{n}\right]&=(m-n)L_{m+n}+\frac{c}{12}(m^3-m)\delta_{m+n,0}
\\
\left[L_{n},c\right]&=0
	\end{align}
\end{subequations}
where $c$ is the \textit{central charge}. A small conformal transformation in radial coordinates is 
\begin{align}
w  \rightarrow w' &= w + \varepsilon(w).
\label{wplusepsilon}
\end{align}
and similarly for $\bar{w}$. The function $\varepsilon$ can be written as Fourier decomposition
	\begin{align}
	\varepsilon &= \sum_n \varepsilon_n w^{-n+1}.
	\end{align}
Since $x'^+$ has to be real as well, we need to demand
\begin{align}
\varepsilon_n^* &= -\varepsilon_{-n}+ \sum\limits_{n_1 = -\infty}^\infty \varepsilon_{-n_1} \cdot \varepsilon_{n_1-n} + \mO(\varepsilon)^3
\label{epsilonunitarity}
\end{align}
which is the extension of the result of \cite{Mandal:2014wfa} to second order. 

\subsection{Conformal transformations}
\label{sec::U}
Under a conformal transformation, the energy-momentum tensor transforms as
	\begin{align}
	\tilde{T}(w')&=\left(\frac{\partial w'}{\partial w}\right)^{-2}\left(T(w)-\frac{c}{12}S(w',w)\right), 
	\end{align}
where $S(w',w)$ is the \textit{Schwarzian derivative}
\begin{align}
S(w',w)=\left(\frac{\partial^3 w'}{\partial w^3}\right)\left(\frac{\partial w'}{\partial w}\right)^{-1}-\frac{3}{2}\left(\frac{\partial^2 w'}{\partial w^2}\right)^2\left(\frac{\partial w'}{\partial w}\right)^{-2}.
\end{align}
Our goal is to explicitly construct an operator $U(\varepsilon)$ that generates this transformation to second order in $\varepsilon$. In order to do this, we calculate the change of the energy-momentum tensor to be 
\begin{subequations}
\begin{align}
\delta T(w)\equiv& \tilde{T}(w)-T(w) 
\label{deltaT}
\\
=&-\varepsilon (w) T'(w)-2 T(w) \varepsilon '(w)-\frac{1}{12} c \varepsilon ^{(3)}(w) \nonumber
\\
&+\frac{1}{12} c \varepsilon ^{(4)}(w) \varepsilon (w)+\frac{1}{8} c \varepsilon ''(w)^2+\frac{1}{4} c \varepsilon ^{(3)}(w) \varepsilon '(w)+\frac{1}{2} \varepsilon (w)^2 T''(w) \nonumber
\\
&+3 \varepsilon (w) T'(w) \varepsilon '(w)+2 T(w) \varepsilon (w) \varepsilon ''(w)+3 T(w) \varepsilon '(w)^2,
\end{align}
\end{subequations}
where the second line is the first order result also found in \cite{Mandal:2014wfa}. At the same time, we have $\delta T(w)\equiv\sum_n \delta L_n w^{-n-2}$ and consequently 
\begin{subequations}
\begin{align}
\delta L_k
&=\Sn  \varepsilon _{n} \left[L_{k},L_{-n}\right] +
\frac{1}{2}\Smn \varepsilon _{n}\varepsilon _{m}\left[L_{-n},\left[L_{-m},L_k\right]\right]
+
\Smn\frac{n+m-2}{4} \varepsilon _{n} \varepsilon _{m}\left[L_k,L_{-n-m}\right] + \mO(\varepsilon)^3,
\label{deltaL1}
\\
&\overset{!}{=} U^\dagger L_k U -L_k. 
\end{align}\label{deltaL}%
\end{subequations}
Defining $\mL\equiv\Sn\e_n L_{-n}$, the first order result of \cite{Mandal:2014wfa} was $U(\varepsilon)=e^{\mL}$ up to terms of order $\mO(\varepsilon^2)$. The second term we obtain in \eqref{deltaL1} also arises from this form of $U$ as is obvious from the formula
$e^X Y e^{-X}=Y+[X,Y]+\frac{1}{2}[X,[X,Y]]+...$\,. 
However, we have to include a correction which produces the third term.
So for the form of $\U$ including terms of order $\mO(\varepsilon^2)$, we make the ansatz
\begin{align}
\U= e^{\mL+ \mB}
\label{U}
\end{align}
where $\mB$ is an operator of $\mO(\e^2)$.  Examining \eqref{deltaL} yields
\begin{align}
\mB=\Smn \frac{m+n-2}{4} \e_m \e_n L_{-n-m}. \label{ResultB}
\end{align}
Let us perform some consistency checks. For $U$ to be unitary, $\mB$ has to satisfy
\begin{align}
\qty(\mL + \mB)^\dagger &= -\mL -\mB \label{UnitarityB} + \mO(\varepsilon)^3,
\end{align}
which \eqref{ResultB} does, using \eqref{epsilonunitarity}. Furthermore, we require that the inverse conformal transformation yields the inverse unitary operator. Calculating the inverse transformation yields
\begin{align}
w &= w' + \varepsilon^{\text{inv.}}(w'), \ \ \   \varepsilon^{\text{inv.}}_n = - \varepsilon_n + \frac{1}{2}\sum_{n_1} \varepsilon_{n_1} \varepsilon_{n-n_1} \qty(2-n) + \mO(\varepsilon)^3.
\end{align}
The subleading terms arise since $\varepsilon^{\text{inv.}}$ is expanded with respect to $w'$ and not $w$. This resulting condition is
\begin{align}
U(\varepsilon^{\text{inv.}}) = U(\varepsilon)^\dagger = e^{-\mL -\mB}. \label{InverseOperator}
\end{align}
$\mB$ is quadratic in $\varepsilon$ and hence identical up to second order for $\varepsilon$ and $\varepsilon^{\text{inv.}}$. Therefore, this consistency condition can also be used to determine $\mB$, yielding the same result as obtained in \eqref{ResultB}.

This determines $\U$ in \eqref{U} up to terms of order $\mO(\e^3)$\footnote{We see that for a small SGD these operators are simple in the sense of \cite{Cottrell:2017ayj}.} \textit{and} up to a possible complex phase shift $\mB\rightarrow\mB+i\beta$ ($\beta\in\mathbb{R}$), as equation \eqref{deltaL1} only depends on commutators of $\mB$. This will play a role in the next section. 

\subsection{Products}
\label{sec::products}

What happens if we do two conformal transformations $\e^{(1,2)}$ after each other? It should be possible to write this as one conformal transformation
\begin{subequations}
\begin{align}
 w' &= w+\e^{(1)}(w) & &
\\
w''&= w'+\e^{(2)}(w') \equiv w+\e^{(3)}(w)
\\[1em]
\e^{(3)}(w)&= \e^{(1)}(w)+\e^{(2)}(w+\e^{(1)}(w)) \nonumber\\
&= \e^{(1)}(w)+\e^{(2)}(w)+\e^{(2)}{}'(w)\e^{(1)}(w)+\mO(\e^3).
\end{align}
\end{subequations}
Consequently, we obtain
\begin{align}
\e^{(3)}_{m}=\e^{(1)}_{m}+\e^{(2)}_{m}+\Sn(1-n)\e^{(1)}_{m-n}\e^{(2)}_{n}.
\end{align}
Using the \textit{Baker-Campbell-Hausdorff} formula
\begin{align}
e^X e^Y =e^Z,\ \ Z=X+Y+\frac{1}{2}[X,Y]+\frac{1}{12}[X,[X,Y]]+...
\end{align}
it is possible to show that indeed 
\begin{align}
U(\e^{(3)})\approx U(\e^{(2)})U(\e^{(1)})
\end{align}
up to terms of order $\mO(\e^3)$ and \textit{up to} an $\e$-dependent complex phase that is proportional to the central charge $c$ and stems from the commutator $[\mL(\e^{(1)}),\mL(\e^{(2)})]$.

There are a few comments about this in order. First of all, in the previous subsection we noted that the requirement that the operators $\U$ in \eqref{U} generate conformal transformations \eqref{deltaT} does not fix an overall complex phase of these operators, so we might define an entire equivalence class of operators (all equal up to a complex phase) and say that each such equivalence class generates the same conformal transformation. 
This means that the operators $U(\e)$ form a  projective representation of the Virasoro group. Acting with such operators on a state like $\left|0\right\rangle$ would then determine the resulting state only up to a complex phase. For \textit{our} purposes, this is sufficient, as we assume that states that differ only by a complex phase have the same complexity \cite{Susskind:2014jwa}.\footnote{I.e.~we assume physical states to be defined modulo overall phase. 
From the holographic perspective, this also makes sense as we could try to take a state with a holographic dual and reconstruct its bulk geometry from data about entanglement entropies of subregions. This reconstruction will not depend on the overall phase, so neither should other geometrical probes that can be calculated from the reconstructed bulk.} Upon inspecting equation \eqref{fidsusdef}, the reader will also notice that fidelity susceptibility would not depend on an overall complex phase of one of the states either.

A more formal treatment of the operators $\U$, taking care of the extra phases introduced by the central charge $c\neq0$ would have to take into account the \textit{Bott-cocycle} as outlined in \cite{Oblak:2017ect}.

\subsection{Real space}
\label{sec::realspace}

Let us return to our previous notation. The conformal transformation induced by the SGD is
\begin{align}
x^+ &=  G_+(\tilde x^+) = \tilde x^+ +   \ep ~g_+(\tilde x^+) 
\end{align}
Comparison with \eqref{wisexp} and \eqref{wplusepsilon} yields the identification
\begin{align}
\exp\left(- \ep ~\frac{2\pi i}{\ell}  \cdot g_+\right) &= 1+\frac{\varepsilon}{w}.
\end{align}
The Fourier decomposition of $g_+$ is
\begin{subequations}
	\begin{align}
	g_+ &= \sum_n g_{+,n} w'^{-n},\\
	g_{+,n}^* &= g_{+,-n}.
	\end{align}
\end{subequations}
Here we have to be careful: The Fourier decomposition of $\varepsilon$ is with respect to $w$, whereas the Fourier decomposition of $g_+$ is with respect to $w'$. 
This produces an additional contribution when expressing $\varepsilon_n$ in terms of $g_{+,n}$. Up to second order in $g_+$, we obtain
\begin{subequations}
	\begin{align}
	\frac{\varepsilon}{w} &\approx - \ep ~\frac{2\pi i}{\ell}  \cdot g_+ - \ep^2 ~\frac{2\pi^2 }{\ell^2}  \cdot g_+^2, \\
	\varepsilon_n &\approx- \ep ~\frac{2\pi i}{\ell}   g_{+,n}+  \ep^2 ~\frac{2(n-1)\pi^2}{\ell^2} \sum_{n_1}  g_{+,n_1}g_{+,n-n_1}.
	\end{align}
\end{subequations}
\Comment[inline,caption={Change from $w'$ to $w$}]{
\begin{minipage}{0.75\linewidth}
	\begin{align*}
	w' &= w \exp(-\ep\frac{2\pi i}{\ell}g_+) \\
	w'^{-n} &\approx  w^{-n}(1+n \ep\frac{2\pi i}{\ell}g_+) \\[1em]
	- \ep ~\frac{2\pi i}{\ell}  \cdot g_+  &= - \ep ~\frac{2\pi i}{\ell}  \cdot \Sn g_{+,n}w'^{-n} \\
	&\approx - \ep ~\frac{2\pi i}{\ell}  \cdot \Sn g_{+,n}w^{-n} + \ep^2 ~\frac{4\pi^2}{\ell^2}  \cdot \Smn n g_{+,n}g_{+,m}w^{-n-m}\\
	&\approx - \ep ~\frac{2\pi i}{\ell}  \cdot \Sn g_{+,n}w^{-n} + \ep^2 ~\frac{2\pi^2}{\ell^2}  \cdot \Smn (n+m) g_{+,n}g_{+,m}w^{-n-m}
	\end{align*}
\end{minipage}
}
Expressing the previous results in terms of $g_+$ yields
\begin{subequations}
	\begin{align}
	\delta L_m &= - \ep ~\frac{2\pi i}{\ell} \sum_{n} g_{+,n} \comm{L_m}{L_{-n}}
	-  \ep^2 ~\frac{2 \pi^2}{\ell^2} \sum_{n_i} g_{+,n_1}g_{+,n_2} \comm{L_{-n_1}}{\comm{L_{-n_2}}{L_{m}}} \nonumber \\
	&~~~+ \ep^2 ~ \frac{\pi^2}{\ell^2} \sum_{n_i} (n_1+n_2) g_{+,n_1}g_{+,n_2} \comm{L_m}{L_{-n_1-n_2}} + \mO(\ep)^3 \\
	U(g_+) &= 1- \ep ~\frac{2\pi i}{\ell} \sum_{n} g_{+,n}L_{-n}
	-  \ep^2~\frac{2 \pi^2}{\ell^2} \sum_{n_i} g_{+,n_1}g_{+,n_2} L_{-n_1}L_{-n_2} \nonumber \\
	&~~~+ \ep^2 ~ \frac{\pi^2}{\ell^2} \sum_{n_i} (n_1+n_2) g_{+,n_1}g_{+,n_2} L_{-n_1-n_2}+ \mO(\ep)^3.
	\end{align}
\end{subequations}
This operator is unitary, keeping in mind that $g_{+,n}^* = g_{+,-n}$. 
Now, let us carefully take the limit of an infinite spatial cycle, i.e.~$\ell \rightarrow \infty$. For that, we define 
\begin{align}
\xi =& \frac{n}{\ell}, \ \ \  d \xi = \frac{1}{\ell}.
\end{align}
The expansions in $w$ are Fourier transformations in this limit\footnote{We see that in this limit, the $\varepsilon_n$ are not suitable any more, as they go as $g_+/\ell = \hat g_+/\ell^2 \rightarrow 0$.}
\begin{subequations}
	\begin{align}\
	g_+(\tilde x^+)=&  \int d\xi  ~\underbrace{\ell\cdot g_{+,\ell \xi}}_{\hat g_+ ( \xi)} \exp (-2 \pi i\cdot \tilde x^+ \cdot \xi),\\
	T_{++}(x^+) =& -4\pi^2   \int d\xi ~\underbrace{\frac{1}{\ell^2}\ell\cdot L_{\ell \xi}}_{\hat L(\xi)} ~ \exp(-2 \pi i\cdot x^+ \cdot \xi).
	\end{align}
\end{subequations}
The operator we look at consequently is
\begin{subequations}
	\begin{align}
	U(g_+) &= 1- \ep ~ 2\pi i \int d\xi ~\hat g_{+}(\xi) \hat L(-\xi)
	-  \ep^2 ~ 2 \pi^2\iint d\xi_1  d\xi_2~ \hat g_{+}(\xi_1)\hat g_{+}(\xi_2) \hat L(-\xi_1)\hat L(-\xi_2) \nonumber \\
	&~~~+  \ep^2 ~\pi^2 \iint d\xi_1 d\xi_2~ (\xi_1+\xi_2) \hat g_{+}(\xi_1)\hat g_{+}(\xi_2) \hat L(-\xi_1-\xi_2)+ \mO(\ep)^3, \\
	&= 1+ \ep ~\frac{i}{2\pi} \int dx~ g_+(x)\cdot T_{++}(x)- \ep^2 ~\frac{1}{8\pi^2} \iint dx_1 dx_2~ g_+(x_1)g_+(x_2)\cdot T_{++}(x_1) T_{++}(x_2) \nonumber \\
&~~~-\ep^2 ~\frac{i}{8\pi} \iint dx_1 dx_2~ g_+'(x_1)g_+(x_2)\cdot T_{++}(x_1+x_2)  + \mO(\ep)^3
	\end{align}
\end{subequations}
and similarly for $U(g_-)$.
